\documentclass[%
reprint,
superscriptaddress,
groupedaddress,
 amsmath,amssymb,
 aps,
prb,
]{revtex4-1}

\usepackage{graphicx}
\usepackage{dcolumn}
\usepackage{bm}
\usepackage{graphicx}
\usepackage{placeins}

\begin{document}

\title{Nonlinear Dynamics and Dissipation of a Curvilinear Vortex Driven by a Strong Time-Dependent Meissner Current.}

\author{W. P. M. R. Pathirana and A. Gurevich.} 
\affiliation{
Department of Physics and Center for Accelerator Science, Old Dominion University, Norfolk, Virginia, USA 
}
	

%
\begin{abstract}
	
We report numerical simulations of large-amplitude oscillations of a trapped vortex line under a strong ac magnetic field $H(t)=H\sin\omega t$ parallel to the surface. The power dissipated by an oscillating vortex segment driven by the surface ac Meissner currents was calculated by taking into account the nonlinear vortex line tension, vortex mass and a nonlinear Larkin-Ovchinnikov (LO) viscous drag coefficient $\eta(v)$. We show that the LO decrease of $\eta(v)$ with the vortex velocity $v$ can radically change the field dependence of the surface resistance $R_i(H)$ caused by trapped vortices. At low frequencies $R_i(H) $ exhibits a conventional increases with $H$, but as $\omega$ increases, the surface resistance becomes a nonmonotonic function of $H$ which {\it decreases} with $H$ at higher fields. The effects of frequency, pin spacing and the mean free path $l_i $ on the field dependence of $R_{i}(H) $ were calculated. It is shown that, as the surface gets dirtier and $l_i$ decreases, the anomalous drop of $ R_{i}(H) $ with $H$ shifts to lower fields which can be much smaller than the lower critical magnetic field.  Our numerical simulations also show that the LO decrease of $\eta(v)$ with $v$  can cause a vortex bending instability at high field amplitudes and frequencies, giving rise to the formation of dynamic kinks along the vortex. Measurements of $R_i(H)$ caused by sparse vortices trapped perpendicular to the surface can offer opportunities to investigate an extreme nonlinear dynamics of vortices driven by strong current densities up to the depairing limit at  low temperatures. The behavior of $R_i(H)$ which can be tuned by varying the rf frequency or concentration of nonmagnetic impurities is not masked by strong heating effects characteristic of dc or pulse transport measurements.  

\end{abstract}
 
 \maketitle

\section{Introduction}

The dynamics of current-driven vortex matter in superconductors is of major importance both for the fundamental vortex physics and for achieving high non-dissipative currents in applications. Materials advances in incorporating artificial pinning centers that immobilize vortices have resulted in critical current densities $J_c$ as high as $10-30 \% $ of the depairing current density $J_d$ at which the superconducting state breaks down \cite{jc1,jc2,jc3,jc4,jc5}. At such high current densities $J$, once a vortex gets depinned from a defect, it can move with very high velocity $v$ and dissipate much power. This phenomenon is critical for many applications, such as high-field magnets \cite{mag1,mag2,mag3}, THz radiation sources \cite{thz1,thz2}, or resonator cavities for particle accelerators \cite{Padam_book,gurevich2012}. Yet, the extreme dynamics of curvilinear elastic vortices driven by very strong currents close to the depairing limit $J\sim J_d$ has not been well understood.  

At high current densities with $J\gg J_c$ the effect of pinning diminishes and the velocity of a vortex $v$ is mainly determined by the balance of the driving Lorentz force $F_L=\phi_0J$ and the viscous drag force, $F_d=\eta(v)v$. At small $v$ the vortex drag coefficient $\eta_0\simeq \phi_0^2/2\pi\xi^2\rho_n$ is independent of $v$, where $\xi$ is the coherence  length, $\rho_n$ is the normal state resistivity, and $\phi_0$ is the magnetic flux quantum. Since the current density cannot exceed the depairing limit $J_d\simeq \phi_0/4\pi\mu_0\lambda^2\xi$ at which the speed of the superconducting condensate reaches the pair-breaking velocity $v_d=\hbar/\pi m\xi$, the maximum vortex velocity can be estimated as $v_c\sim\phi_0J_d/\eta_0= \rho_n\xi/2\mu_0\lambda^2$, where $\lambda$ is the magnetic penetration depth and $m$ is the effective electron mass. For instance, for clean Nb with $\xi\simeq \lambda\simeq 40$ nm and $\rho_n\simeq 1 $~n$\Omega\cdot$m, we have $v_d\simeq 0.9$ km/s and $v_c\simeq 10$ km/s, that is, the vortex can move faster than than the maximum drift velocity of the condensate. Vortices moving much faster than current superflow which drives them have been observed by scanning SQUID on tip microscopy in dirty Pb films in which $v_c$ and $v_d$ can differ by two orders of magnitude ~\cite{embon}. 

At high velocities the vortex drag coefficient $\eta$ is determined by complex nonequilibrium processes in the vortex core \cite{kramer,kopnin} elongated along the direction motion, as was shown by simulations of the time-dependent Ginzburg-Landau (TDGL) equations  \cite{embon,tdgl1,tdgl2,anl}. As a result, $\eta(v)$ becomes essentially dependent on $v$. For instance, Larkin and Ovchinnikov (LO) have shown that $\eta(v)$ decreases with $v$ because quasiparticles in the core diffuse away from the core at high velocities \cite{LO}, giving:
\begin{equation} 
\eta=\frac{\eta_0 }{1+v^2/v_0^2}.
\label{LO}
\end{equation}
Here the critical LO velocity $v_0 \sim (D/\tau_\epsilon)^{1/2}(1-T/T_c)^{1/4}$ depends on the energy relaxation time $\tau_\epsilon$, where $D$ is the electron diffusivity~ \cite{LO}.  A similar velocity dependence of $\eta(v)$ could also occur due to electron overheating in the moving vortex \cite{shklovsk,kunchur,gc} which would be particularly pronounced at low temperatures.

The LO mechanism predicts a nonmonotonic velocity dependence of the drag force $F_d=\eta(v)v$ which reaches maximum at $v=v_0$, so that $F_d(v)$ can balance the Lorentz forces $F_L=\phi_0J$ only if $v<v_0$.  At $v>v_0$ the velocity of a straight vortex driven by a uniform current density jumps to greater values corresponding to highly dissipative states \cite{LO}. Such  LO instability has been observed on many superconductors \cite{mus1980,klein1985,armenio2007,villard2003,grimaldi2008,inst1,inst2,inst3,inst4} near $T_c$ with typical values of $v_0\sim 0.1-1$ km/s. This instability results in negative differential resistivity and hysteretic jumps on I-V curves. Observation of the LO instability in dc transport measurements in which vortex structures move with high velocities $v\sim v_0$ in thin films at low temperatures is masked by strong heating effects \cite{tdgl2,inst3,inst4,gm}.  

The extreme dynamics of vortices driven by strong currents at low temperatures can be investigated under conditions in which heating effects are greatly reduced. This geometry is shown in Fig. 1 where a perpendicular vortex trapped in a superconducting slab of thickness $d\gg \lambda$ is exposed to a RF parallel magnetic field $H(t)=H\sin\omega t$ with $\hbar\omega\ll \Delta$. Here sparse trapped vortices are driven by non-dissipative Meissner current flowing in a thin layer $\sim\lambda$ at the surface so that the net power generated by superfast vortices is much smaller than in pulse flux flow measurements \cite{inst3,kunchur}. It is this situation which is characteristic of superconducting resonant cavities with extremely high quality factors $Q=R_0/R_s\sim 10^{10}- 10^{11} $ controlled by a very small surface resistance $R_s\approx 10-30$ n$\Omega$ for Nb at $2$ K and $1-2$ GHz~\cite{Padam_book,gurevich2012}. Part of this $R_s$ comes from the exponentially small quasiparticle BCS contribution $R_{BCS}\propto\omega^2\exp(-\Delta/T)$ but another one is a weakly-temperature dependent residual resistance $R_i$ which can account for $ \gtrsim 20\%$ of $R_s$ in Nb ~\cite{Padam_book} and $ \gtrsim 50\%$ for Nb\textsubscript{3}Sn at 2K ~\cite{liarte}. Much of this contribution  to $R_i$ comes from trapped vortices generated during the cavity cool down through $T_c$ at which the lower critical field $H_{c1}(T)$ vanishes ~\cite{tf1,tf2,tf3,tf4,tf5,tf6,tf7,tf8}. In this case even small stray fields $ H>H_{c1}(T) $ such as a few $\%$ of the earth magnetic field can produce vortices in the cavity. During the subsequent cooldown to $T\simeq 2$ K some of these vortices exit but some get trapped by material defects and produce hotspots caused by oscillating vortex segments depicted in Fig. 1. Superconducting cavities can thus provide a unique testbed for the investigation of extreme dynamics of vortices driven by strong Meissner currents which can reach $J\sim J_d$ at the surface at $B=\mu_0H\simeq \mu_0H_c$, where $H_c$ is a thermodynamic critical field. This has indeed been achieved on Nb cavities at $1-2$ GHz and  $2$K at $B\simeq 200-230$ mT. Here the high quality factors $Q\sim 10^{10}-10^{11}$ indicate very low dissipated power in sparse vortex bundles trapped in the cavity ~\cite{tf2}.  

The nonlinear dynamics of the vortex driven by strong Meissner currents depicted in Fig. 1 brings about the following issues. 1. The estimates presented above show that the velocity of the vortex tip at the surface $v(0,t)\sim v_d$ at $H\simeq H_c$ can exceed the LO critical velocity $v_0$. However, once $v(0,t)$ exceeds $v_0$, the LO jump-wise instability does not occur because the tip is connected to the rest of the elastic vortex line. The questions are then what happens if a part of an elastic vortex moves faster than the LO critical velocity while the rest of the vortex does not, and  if so, can the LO dynamic instability manifest itself in a shape instability of a distorted vortex? 2. What are the dependencies of the power dissipated by an elastic curvilinear vortex driven by a surface Meissner current on the field amplitude and frequency? 3. To what extent can the nonlinear surface resistance of trapped vortices be tuned by varying the concentration of nonmagnetic impurities?   Addressing these issues is the goal of this work.

The surface impedance of the mixed state of vortices parallel to the surface under weak RF field has been extensively studied in the literature~ \cite{ce,blatter,ehb}. Low-field power losses generated by flexible pinned vortex segments both parallel and perpendicular to the surface were calculated in Refs. \onlinecite{gc,tf2}. Nonlinear quasi-static electromagnetic response of perpendicular vortices has been recently addressed both in the limit of weak collective pinning  ~\cite{liarte} and the strong pinning limit ~\cite{dima} in which the depinning of elastic vortices under the Lorentz force becomes hysteretic ~\cite{labusch}. However, the extreme nonlinear dynamics of a long elastic vortex under a strong rf surface current has not been addressed. In this work we calculate the power $P(H)=R_i(H)H^2/2$ generated by such oscillating vortex and the corresponding surface resistance $R_i(H)$ as functions of the field amplitude $H$, frequency, the mean free path and the location of the pining center from the surface, taking into account the nonlinear elasticity of the distorted vortex, nonlinear viscous drag force, and the vortex mass. We show that the LO velocity dependence of the vortex viscosity (\ref{LO}) results in an anomalous {\it decrease} of $R_i(H)$ with $H$ at strong RF field.  A  significant decrease of $R_i(H)$ with the RF field extending up to $H\simeq (0.3-0.5)H_c$ has been indeed observed by several groups on alloyed Nb cavities ~\cite{cav1,cav2,cav3,cav4,cav5}. One mechanism to this effect can come from the nonlinear response of quasiparticle in the Meissner state ~\cite{ag_prl,ag_sust,kg}. In this work we show that sparse trapped vortices can provide another mechanism of the decrease of $R_i(H)$ with $H$ which becomes more pronounced as the frequency increases. We also show that the LO nonlinear drag force can result in a shape instability and the formation of dynamic kinks along the elastic fortes at large fields and frequencies.   

The paper is organized as follows. Sec. II introduces the main nonlinear dynamic equation for a current driven elastic vortex perpendicular to the surface and defines the essential control parameters. In Sec. III we present an analytical solution that describes a vortex at low-frequencies and show that $R_i(H)$ does decrease with $H$ above a crossover frequency. Sec. IV contains the results of our extensive numerical simulations of the dynamic equation for the oscillating vortex at arbitrary frequencies and calculate $R_i(H)$ as functions of $H$, frequency and the mean free path. In Sec. V we demonstrate  a  shape instability of a strongly driven elastic vortex with the LO nonlinear viscous drag. Sec. VI presents the discussion and broader implications of our results.    

\section{Dynamic equations.} \label{DE}

\begin{figure}[!htb]
	\centering
	\includegraphics[scale=0.48]{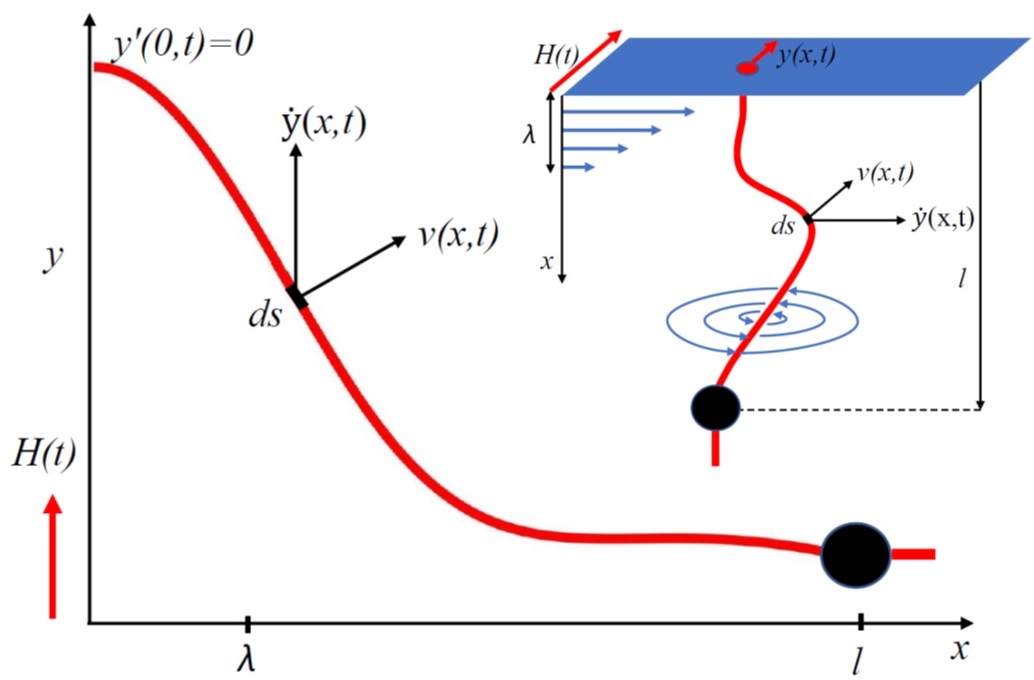}
	\caption{\label{fig:Fig1} A curvilinear vortex driven by the rf surface current. The black dot shows a pinning center, for example, a nonsuperconducting precipitate. Inset shows the geometry of a perpendicular vortex trapped in a superconducting slab exposed to a parallel magnetic field $H(t)=H\sin\omega t$. The blue arrows depict supecurrents circulating around the vortex and Meissner screening currents at the surface.} 
\end{figure}

Consider a single vortex pinned by a materials defect as shown in the Figure \ref{fig:Fig1}. 
The motion of a vortex is determined by its vertical displacement $y(x,t) $ as a function of $x$ and $t$, where the tip of the vortex is perpendicular to the surface \cite{ehb} so that $y'(0)=0 $. The equation for $y(x,t)$ is obtained using the equation for the local velocity $v(x,t)$ normal to the curvilinear vortex:
\begin{equation} 
M \dot{v}+\eta (v) v=\frac{\epsilon}{R} - \frac{\phi_0 H}{\lambda}e^{-z/\lambda}\sin (2\pi f t),
\label{eq1}
\end{equation}
where $H$ is the amplitude of the applied magnetic field $H\sin\omega t$ with the frequency $f=\omega/2\pi$ , $\lambda$ is the London penetration depth, $M$ is the vortex mass per unit length, $\epsilon=\phi_0^2(\ln\kappa+0.5)/4\pi\mu_0\lambda^2$ is the vortex line energy, $ \kappa=\lambda/\xi $ is the Ginzburg-Landau (GL) parameter, $\xi$ is the coherence length, $R^{-1}$ is the local curvature of the vortex line, and the overdot means a time derivative.

Equation (\ref{eq1}) represents a balance of local forces acting perpendicular to a curvilinear vortex: the inertial and viscous drag forces in the left hand side are balanced by the elastic and Lorentz forces in the right hand side.  Here it is assumed that: 1. The motion of a magnetic  vortex in a type-II superconductor is described by the London model, the dynamics of the vortex core is incorporated in $\eta(v)$ and $M$. 2. Meissner current densities are not very close to $J_d$ so pairbreaking effects are negligible and the London model is applicable. 3. The Magnus force causing a small Hall angle \cite{v1,v2,v3} is negligible and the velocity ${\bf v}(x,t)$ only has components $v_x(x,t)$ and $v_y(x,t)$ perpendicular to the rf current flowing along the $z$ axis. 4. The low frequency rf field ($\hbar\omega\ll \Delta$) does not produce quasiparticles, and the quasi-static London equations are applicable \cite{tinkham}. 5. Spatial distortions of a vortex occur over large scales $\gtrsim\lambda$ for which the elastic nonlocality of $\epsilon$ (see, e.g., Refs. \onlinecite{blatter,ehb})  can be neglected. The effect of nonlocality of $\epsilon$ on the power dissipated by the vortex under a weak rf field was considered in Ref. \onlinecite{tf2}.   

The local perpendicular velocity $v(x,t)$ of a small vortex segment $ds$ in Eq. (\ref{eq1}) is related to the velocity $\dot{y}(x,t)$ of the vortex line at a fixed point $x$ by:
\begin{equation}
 v(x,t)=\frac{\dot{y}(x,t)}{\sqrt{1+y'^2}}.
 \label{par}
 \end{equation}
This relation reflects the fact that each small segment of an overdamped vortex moves along the local normal to the curvilinear vortex line under the action of local perpendicular forces defined by Eq. (\ref{eq1}).    
The term $\epsilon/R$ in Eq. (\ref{eq1}) accounts for a nonlinear elasticity of a vortex in the London model \cite{ehb}, where the local curvature $R^{-1} =y''(1+y'^2)^{-3/2}$ depends on the shape of the vortex line and the prime denotes a partial derivative with respect to $x$. Equations (\ref{eq1}) and (\ref{par}) give the following dimensionless nonlinear partial differential equation for the local displacement $u(x,t)$ of the vortex along the $y-$axis:
\begin{gather} 
\mu\frac{\partial}{\partial t}\left[\frac{\dot{u}}{\sqrt{1+u'^2}}\right]+\frac{\gamma \dot{u}\sqrt{1+{u'}^2}}{1+u'^2+\alpha \dot{{u}}^2}=
\nonumber \\
\frac{u''}{(1+{u'}^2)^{3/2}}-\beta_t e^{-x},
\label{dyneq} \\
u'(0,t)=0,\qquad u(l,t)=0.
\label{bc0}
\end{gather}
Here $u(x,t)=y(x,t)/\lambda$ is the dimensionless displacement of the vortex along $y$, the coordinate $x$ and time $t$ are in units of $\lambda$ and the rf period, respectively. 
The second boundary condition in Eq. (\ref{bc0}) describes a vortex pinned by a strong defect, but the numerical results presented below are, in fact, not very sensitive to the elementary pinning force of the defect, as elaborated in Sec. \ref{Results}.  The parameters in Eq. (\ref{dyneq}) are given by:
\begin{gather}
\gamma=f/f_0,  \qquad f_0=H_{c1}\rho_n/{H_{c2}\lambda^2 \mu_0},  
\label{gamma}\\
\alpha=\alpha_0\gamma^2, \qquad \alpha_0=(\lambda f_0/v_0)^2,
\label{alpha}\\
\beta_t=\beta \sin(2\pi t), \qquad \beta=H/H_{c1},
\label{beta}\\
\mu=\mu_1 \gamma^2, \qquad \mu_1=\lambda^2f_0^2 M/\phi_0 H_{c1}.
\label{mu}
\end{gather}
Here $H_{c1}=(\phi_0/4\pi\mu_0\lambda^2)(\ln\kappa+0.5)$ and $H_{c2}=\phi_0/2\pi\mu_0\xi^2$ are the lower and upper critical fields, respectively.

The main contribution to the vortex mass in Eq. (\ref{eq1}) comes from quasiparticles in the vortex core ~\cite{kopnin}. The first estimate of $M$ by Suhl ~\cite{suhl} gave $M_s\simeq 2mk_F/\pi^3$, where $m$ is the electron mass, $k_F=(3\pi^2 n_0)^{1/3} $ is the Fermi wave vector and $n_0$ is the electron density. Other contributions which can increase the vortex mass well above $M_s$ have been proposed in the literature ~\cite{vm1,vm2,vm3,vm4}. Measurements of $M$ in Nb ~ \cite{golubchik} gave $M$ some 2 orders of magnitude higher than $M_s$ near $T_c$. In our simulations we assumed that $M$ in Eqs. (\ref{eq1}) and (\ref{dyneq}) is independent of $v$. 

We first estimate characteristic values of $\alpha$, $\gamma$ and $\mu$ for a dirty Nb with $\rho_n\approx3$ n$\Omega\cdot$m, $\lambda=80$ nm, $\xi=20$ nm $\kappa=4$, $v_0=0.1$ km/s, $k_F=1.2\cdot 10^{10}$ m$^{-1}$ (see, e.g., Ref. \onlinecite{ashcroft}), and taking $M=80M_s=5.6\cdot 10^{-20}$ kg/m. Hence, $f_0\simeq 22$ GHz, $\alpha_0\simeq 309$, and $\mu_1\simeq 0.0022$, so that $\gamma\simeq 0.045$, $\mu\simeq 4.5\cdot 10^{-6}$, and $\alpha\simeq 0.64$ at $f=1$ GHz. Next we consider  Nb\textsubscript{3}Sn with $\rho_n\approx 1$ $\mu\Omega\cdot$m, $\lambda=111$ nm, $\xi=4.2$ nm, $\kappa=26.4$ ~\cite{liarte}, $v_0=0.1$ km/s, $k_F=6.6\cdot 10^{9}$ m$^{-1}$ (see, e.g., Ref. \onlinecite{ashcroft}), and taking $M=80M_s=3.1\cdot 10^{-20}$ kg/m. Hence, $f_0\simeq 175$ GHz, $\alpha_0\simeq 3.7\cdot 10^{4}$, and $\mu_1\simeq 0.14$, so that $\gamma\simeq 0.006$, $\mu\simeq 4.6\cdot 10^{-6}$, and $\alpha\simeq 1.2$ at $f=1$ GHz. In this frequency range  the dynamic terms in the l.h.s. of Eq. (\ref{dyneq}) are proportional to the small parameters $\gamma$ and $\mu$, and the effect of the vortex mass at $v\ll v_0$ is much weaker than the viscous drag. However, the effect of the dynamic terms increases strongly as the frequency and the field amplitude increase and/or the material becomes dirtier and the mean free path $l_i$ decreases. For instance, in the dirty limit $l_i\lesssim \xi_0$, we have $\lambda\simeq \lambda_0 (\xi_0/l_i)^{1/2}$ and $\xi\simeq (\xi_0 l_i)^{1/2}$, where the subscript $0$ refers to the clean limit values of the parameters.  Thus, $f_0^{dirty}\simeq (l_i/\xi_0)^2f_0^{clean}$, so the parameter $\gamma^{dirty}\simeq  (\xi_0/l_i)^2\gamma^{clean}$ at a given frequency can increase substantially as $l_i$ decreases. 

Another essential parameter is the decay length $L_\omega$ of oscillating bending disturbance along the vortex line induced by a weak RF current at the surface \cite{tf2}
\begin{equation}
L_\omega=\sqrt{\frac{\epsilon}{\eta\omega}}=\frac{\xi}{2\lambda}\sqrt{\frac{g\rho_n}{\pi\mu_0f}}=\frac{\lambda}{\sqrt{2\pi\gamma}},
\label{Lom}
\end{equation}
where $g=\ln(\lambda/\xi)+1/2$, and the vortex mass is neglected. For the above materials parameters of Nb$_3$Sn, we have $L_\omega\simeq 5.15\lambda=572$ nm at 1 GHz. In this case dissipative oscillations of the elastic vortex extend well beyond the rf field penetration depth. Here $L_\omega$ is practically independent of $T$ and decreases as the m.f.p. decreases, $L_\omega ^{dirty}\simeq  L_\omega^{clean}(l_i/\xi_0)^{1/2}$.  Although Eq. (\ref{Lom}) is only applicable to small-amplitude vortex oscillations, the dependence of $L_\omega$ on $\eta$ suggests that the elastic ripple length $L_\omega$ would increase with the RF field, as the velocity of the vortex tip increases and the LO vortex drag diminishes. This qualitative assertion is in agreement with the numerical results presented below.    

Solving Eq. (\ref{dyneq}) for $u(x,t)$ we calculate the power $ P=t_m^{-1}\int_{0}^{t_m}\int_{0}^{l} \eta v^2dxdt $ produced by the drag force along the oscillating vortex and averaged over the time period $t_m$.  It is convenient to define the dimensionless power $p=P/P_0$ and the surface resistance $r_i$ per vortex as follows
\begin{gather} 
p=\gamma^2 \int_{0}^{1}dt\int_{0}^{l}\frac{\dot {u}^2 (1+u'^2)^{1/2}dx}{1+u'^2+\alpha\dot {u}^2},
\label{p} \\
r_i(\beta)=2p(\beta)/\beta^2,
\label{rs}
\end{gather}
where $P_0=\lambda^3f_0^2\eta_0$.  If sparse trapped vortices have the areal density $n_\square=B_0/\phi_0$ corresponding to a small induction $B_0\ll B_{c1}$, the dimensionless $R_i$ is related to the observed surface resistance by $R_i=P_0r_i n_\square/H_{c1}^2$. Using here $f_0$ from Eq. (\ref{gamma}), $\eta_0=\phi_0B_{c2}/\rho_n$, and $B_{c2}=\phi_0/2\pi\xi^2$, we obtain:
\begin{equation}
R_i=\frac{\rho_nB_0}{\lambda B_{c2}} r_i.
\label{rii}
\end{equation}

\section{Low frequency limit}\label{lf}

In this Section we show how the LO velocity dependence of $\eta(v)$ results in a decrease of $R_i(H)$ with the rf field at low frequencies $\gamma\ll 1$. In this case $l\ll L_\omega$ and $u(x,t)$ can be obtained analytically by solving Eq. (\ref{dyneq}) in which all dynamic terms in the left hand side are neglected.  Then integration 
of Eq. (\ref{dyneq}) with $\dot{u}\to 0$ and the boundary condition $u'(0,t)=0$ at the surface gives:
\begin{equation}
\frac{u'}{\sqrt{1+u'^{2}}}=\beta_t(1-e^{-x}).
\label{up}
\end{equation}
Equation (\ref{up}) has a solution only if $u'(l)<\tan\theta$, where the depinning angle $\theta$ quantifies the strength of the pinning center~\cite{ehb,labusch}. The condition $u'(l)=\tan\theta$ thus defines a critical value of the current driving parameter $\beta_c$ above which the vortex segment can no longer be pinned:
\begin{equation}
\beta_c=\frac{\sin\theta}{1-e^{-l}}.
\label{betc}
\end{equation} 
In the strong pinning limit $(\theta\to\pi/2)$, we have $\beta_c=1$ at $l\gg1$. Here $\beta_c(l)$ increases as $l$ decreases, reducing to  $\beta_c\simeq\sin\theta/l$ at $l\ll 1$ for a uniform current ~\cite{ehb}. Integration of Eq. (\ref{up}) with the boundary condition $u(l)=0$ yields a cumbersome formula for $u(x,t)$ which is then used to obtain an analytical formula for a quasi-stationary $\dot{u}(x,t)$ at $\gamma\ll 1$, as described in Appendix \ref{Ap}.  

From Eq. (\ref{up}), it follows that $1+u'^{2}=[1-s(x)^{2}]^{-1}$, where $s(x)=\beta_t(1-e^{-x})$. Then Eq. (\ref{p}) at $\gamma\ll 1$ can be written in the form:
\begin{equation}
p=\gamma^2\int_{0}^{1}dt\int_{0}^{l}\frac{\dot{u}^{2}\sqrt{1-s^2(x)}dx}{1+\alpha_0\gamma^2[1-s^2(x)]\dot{u}^{2}},
\label{sp}
\end{equation}
where $\dot{u}(x,t)$ in the first order in $\dot{\beta}_t$ is given by Eq. (\ref{du}). 

Using Eqs. (\ref{du}) and (\ref{sp})  the field-dependent nonlinear surface resistance $r_i(\beta,f)=2p/\beta^2$ can be  calculated. 
Here the LO factor $\alpha=\alpha_0\gamma^2$ in the denominator changes the behavior of $r_i(\beta,f)$ at $\alpha\gtrsim 1$, which can happen even at $\gamma^2\ll 1$ if $\alpha_0\gg 1$.  Indeed, at $\alpha\ll 1$ Eq. (\ref{sp}) yields the conventional vortex viscous power $p\propto \beta^2\gamma^2$ at low fields and frequencies \cite{gc}. However, in the limit of $\alpha\beta^2\gg 1$, the term $\dot{u}^2$ in Eq. (\ref{sp}) cancels out and $p$ becomes independent of frequency. In this case $p$ and $r_i$ was  calculated analytically in Appendix \ref{Ap}, where Eq. (\ref{sp1}) at $\alpha^{-1/2}\ll \beta\ll 1$ simplifies to:
\begin{equation}
p=\frac{l}{\alpha_0}, \qquad r_i=\frac{2l}{\alpha_0\beta^2}.
\label{univ}
\end{equation}
These results also readily follow from Eqs. (\ref{p}) and (\ref{rs}) in the limit of $\alpha\dot{u}^2\gg 1$ but $u'^2\ll 1$. Here both $p$ and $r_i$ at $\alpha\beta^2\gg 1$ are independent of frequency, whereas the surface resistance $r_i(\beta)$ {\it decreases} with the field amplitude. Obviously, Eq. (\ref{univ}) is no longer applicable at very low field amplitudes $\alpha\beta^2\lesssim 1$ as  $\alpha\dot{u}^2$ in the denominator of Eq. (\ref{p}) becomes negligible. 

\begin{figure}[!htb]
	\includegraphics[scale=0.13]{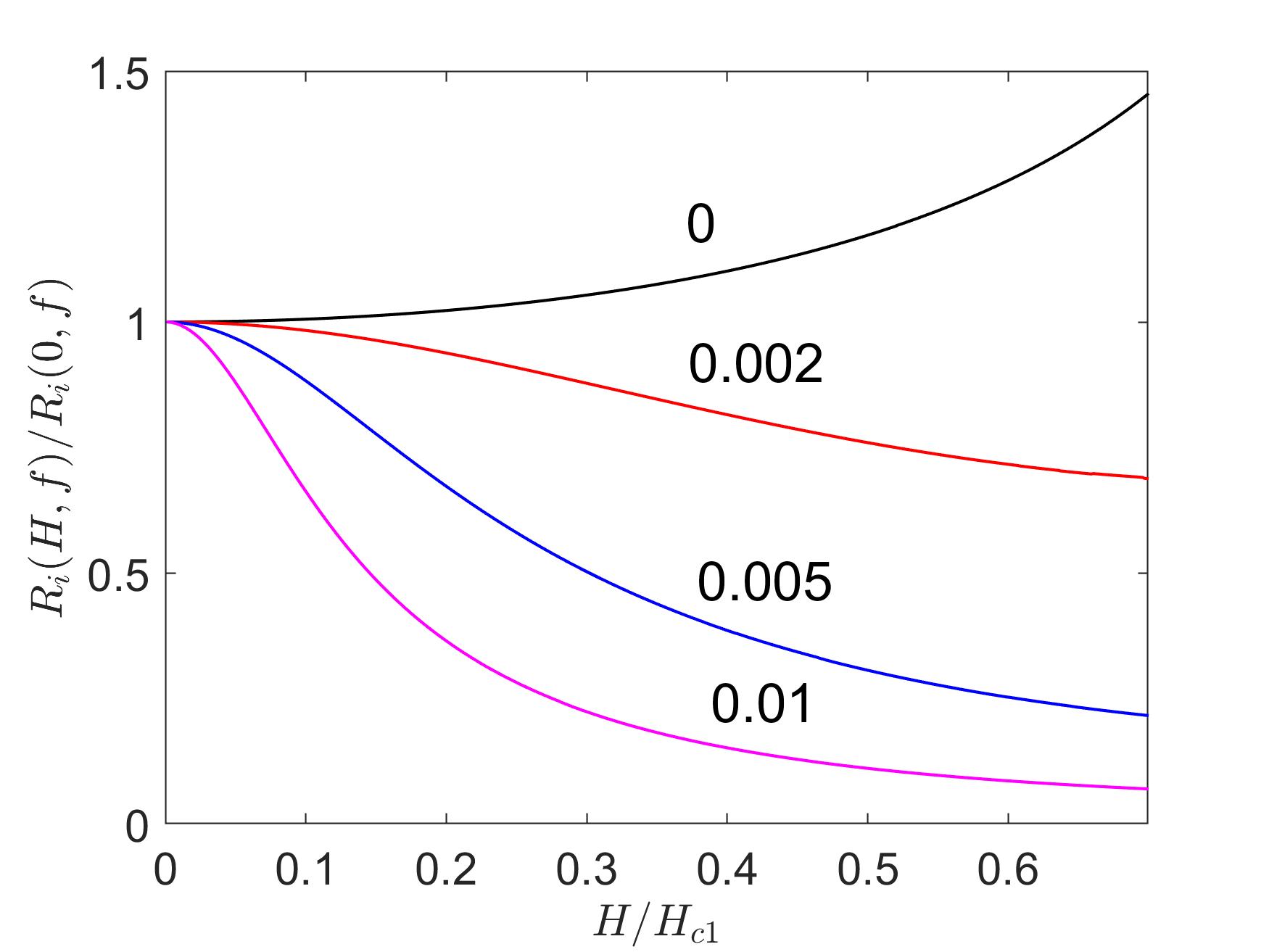}
	\caption{The field-dependent surface resistance $R_i(H)$ calculated from Eq. (\ref{sp}) at $\alpha_0=3\cdot 10^3$, $l=4\lambda$, and the dimensionless frequencies $\gamma$: 0, 0.002, 0.005, and 0.01. }
	\label{fig:Fig2}
\end{figure}
\begin{figure}[!htb]
	\includegraphics[scale=0.13]{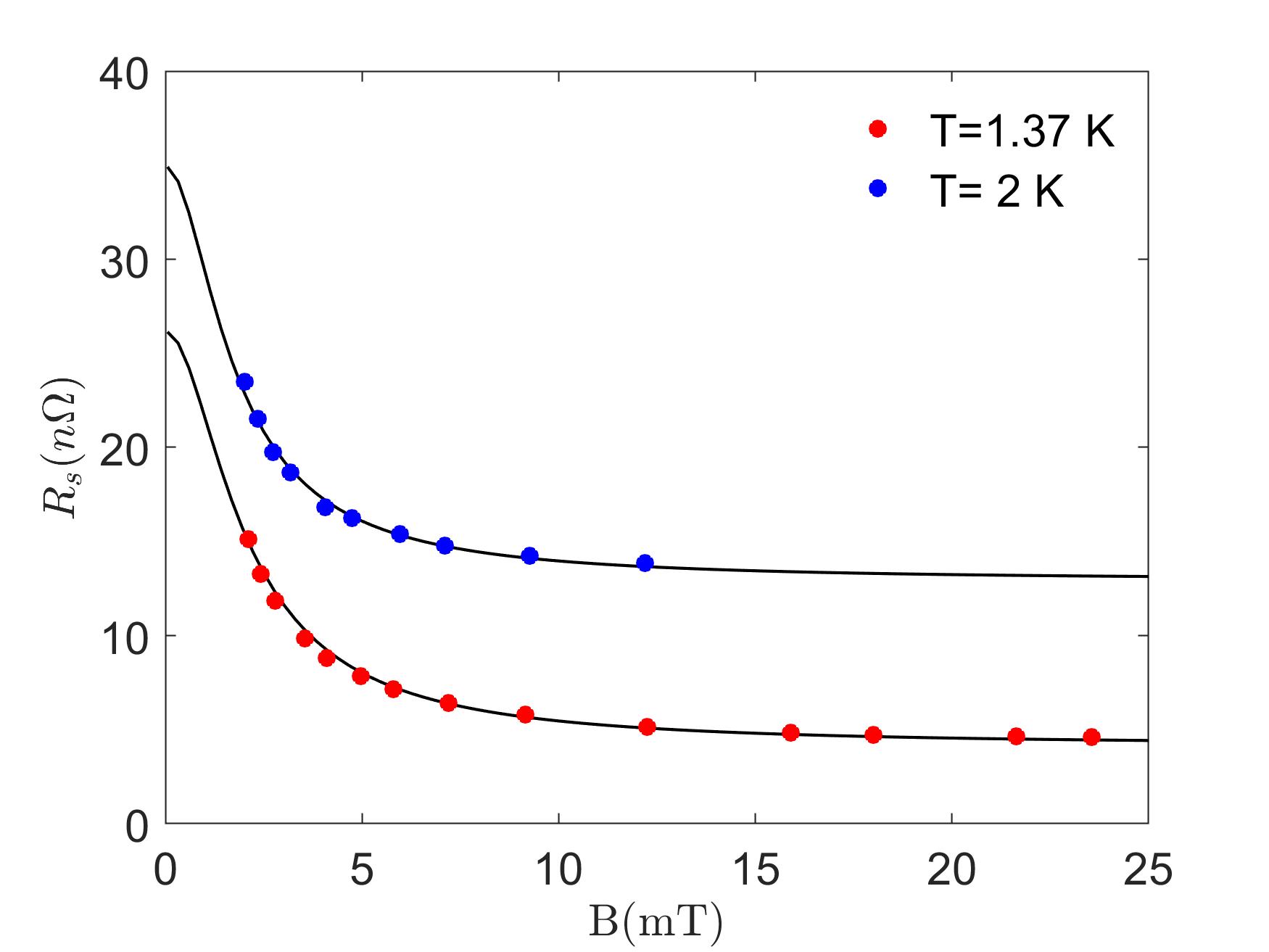}
	\caption{Fits of $R_s(B)=R_i(B)+R_{BCS}$ calculated from Eq. (\ref{sp}) (lines) to the experimental data of Ref. \onlinecite{gigi} (dots) for Nb cavity at $1.467$ GHz, $\gamma=0.052$ and $n_\square=3.67\cdot 10^8$ m$^{-2}$: (a) $\alpha_0 = 3326$ at $T=1.37$ K, (b)  $\alpha_0 = 4380$ at $T=2$ K.}
	\label{fig:Fig3}
\end{figure} 

We calculated the full field dependence $r_i(\beta)$ numerically using Eq. (\ref{sp}), where  $\dot{u}$ is given by Eq. (\ref{du}). These $r_i(\beta)$ curves calculated at different frequencies $\gamma$ are shown in Fig. \ref{fig:Fig2}.  As $\gamma$ increases the LO decrease of $\eta(v)$ with the vortex velocity radically changes the field dependence of $r_i(\beta)$ from an ascending $r_i(\beta)$ at low frequencies to a descending $r_i(\beta)$ at higher frequencies.

The inverse field dependence $R_s\propto H^{-2}$ given by Eq. (\ref{univ}) was observed on Nb cavities subject to a mild heat treatment~ \cite{gigi}.
As an illustration, Fig. \ref{fig:Fig3} shows the fits of Eqs. (\ref{du}) and (\ref{sp}) to $R_s(B)$ measured at $T=1.37$ K and $T=2$ K on a 1.467 GHz single-cell cavity ~\cite{gigi}.  The fit is done for a moderately dirty Nb with  $\rho_n = 2.1$ n$\Omega\cdot$m, $\lambda=70.2$ nm, $\xi=22.8$ nm and $l=3\lambda$, in which case 1.467 GHz corresponds to $\gamma = 0.052$. Here the surface resistance was taken in the form $R_{s}(B)=R_i(B)+ R_{BCS}$,  where $R_{BCS}(T)$ is a background BCS resistance, and $R_i$ is given by Eqs. (\ref{rii}) and (\ref{sp}). It is assumed that a mean areal density of vortices $n_\square=B_0/\phi_0$ was trapped in the superconductor during the cavity cooldown through $T_c$. The fit is then performed at the fixed $\gamma$, regarding $\alpha_0(T)$ and $R_{BCS}(T)$ as independent adjustable parameters at 1.37 K and 2 K, and $n_\square$ as another fit parameter limited by the condition $n_\square(1.37K)=n_\square(2K)$ that the measurements were done on the same cavity.

The fits shown in Fig. \ref{fig:Fig3} are obtained for $R_{BCS}=4.2$ n$\Omega$ and $\alpha_0\approx 3326$ corresponding to $v_0 = 35\,$m/s at $T=1.37$ K, and 
$R_{BCS}=13$ n$\Omega$ and $\alpha_0 = 4380$ corresponding to $v_0 = 30.1\,$m/s at $T=2$ K. 
The fit also gives the mean flux density $n_\square = 3.67\cdot10^8$ m$ ^{-2} $ which translates to the trapped field $B_0=\phi_0n_\square\simeq 0.73~\mu$T 
much smaller than the Earth field $B_E\simeq 20-60~\mu$T. This is consistent with the fact that the cavities measured in Ref. \onlinecite{gigi} were 
magnetically screened during the cooldown through $T_c$ so that the residual field was reduced to     
$B_0\sim 10^{-2}B_E$.  Because trapped flux in Nb cavities is usually localized in bundles of sparse vortices pinned by randomly distributed materials defects~ \cite{tf2}, the observed  $\langle R_i\rangle$ results from averaging over the local values of $B_0(\textbf{r})$ and pin spacings $l$ from the surface ~\cite{ag_sust}.  

\section{Numerical Results} \label{Results}

The results presented above show that the LO velocity dependence of $\eta(v)$ can produce an anomalous decrease of $R_i(H)$ with $H$ as the frequency increases. This brings about several points which can be essential for experimental investigations of this effect:  1. The field dependence of $R_i(H)$ in a broader frequency range in which the quasi-static approximation of Sec. \ref{lf} is no longer applicable, 2. The effect of the length of the pined vortex segment $l$ on $R_i(H)$, 3. The effect of nonmagnetic impurities on $R_i(H)$ which can be used to tune the field and frequency dependencies of $R_i(H,f)$ by alloying the surface of a superconductor.  Addressing these issues require numerical simulations of the nonlinear dynamics of a vortex at arbitrary field amplitudes and frequencies.

In this section we present results of numerical solution of Eqs. (\ref{dyneq}) and (\ref{bc0}) using COMSOL \cite{comsol}. Here the boundary condition $u(l,t)=0$ corresponds to a strong pin with $\theta=\pi/2$ in Eq. (\ref{betc}), but the results for a long vortex segment $l\gg \lambda$ are, in fact, independent of the elementary pinning force $f_p(y)$. Indeed,  if $l\gg L_\omega$ bending oscillations along the vortex do not reach the pin so $u(l,t)=0$ is basically satisfied for any $f_p(y)$.  Yet even if $l<L_\omega$, the details of $f_p(y)$ have a little effect on $R_i(H)$ because incorporating a realistic $f_p(y)$ for the core pinning~\cite{ce} in Eq. (\ref{dyneq})  accounts for small-amplitude ($y(l,t)\lesssim  \xi$) oscillations of the vortex core at the pin, as opposed to the condition of a fixed vortex core, $y(l,t)=0$. Since $R_i(H)$ is determined by large-amplitude swings of a vortex segment between the pin and the surface, taking into account small oscillations of $y(l,t)$ only gives a small correction to $R_i(H)$. For this reason we used the simple boundary condition $u(l,t)=0$. 
  
For the parameters of Nb$_3$Sn mentioned above, we have $\gamma\approx 0.006$, $0.06$ and $0.6$ at 1 GHz, 10 GHz and 100 GHz, respectively. Given the lack of experimental data on $v_0(T)$ for Nb\textsubscript{3}Sn and other superconductors at $T\ll T_c$, we solved Eq. (\ref{dyneq}) numerically for different values of $\gamma$ and $\alpha=\alpha_0\gamma^2=0,1,10$ and $100$.  The corresponding values of $v_0$ cover the typical $v_0\sim 0.1-1$ km/s near $T_c$ ~\cite{grimaldi2008} and take into account the observed decrease of $v_0\propto [D/\tau_{\epsilon}]^{1/2}$ at low temperatures ~\cite{inst1} where the time of energy relaxation on phonons $\tau_\epsilon(T)$ increases as $T$ decreases \cite{kopnin}. 

\begin{figure}[!htb]
	\centering
	\includegraphics[width=\columnwidth]{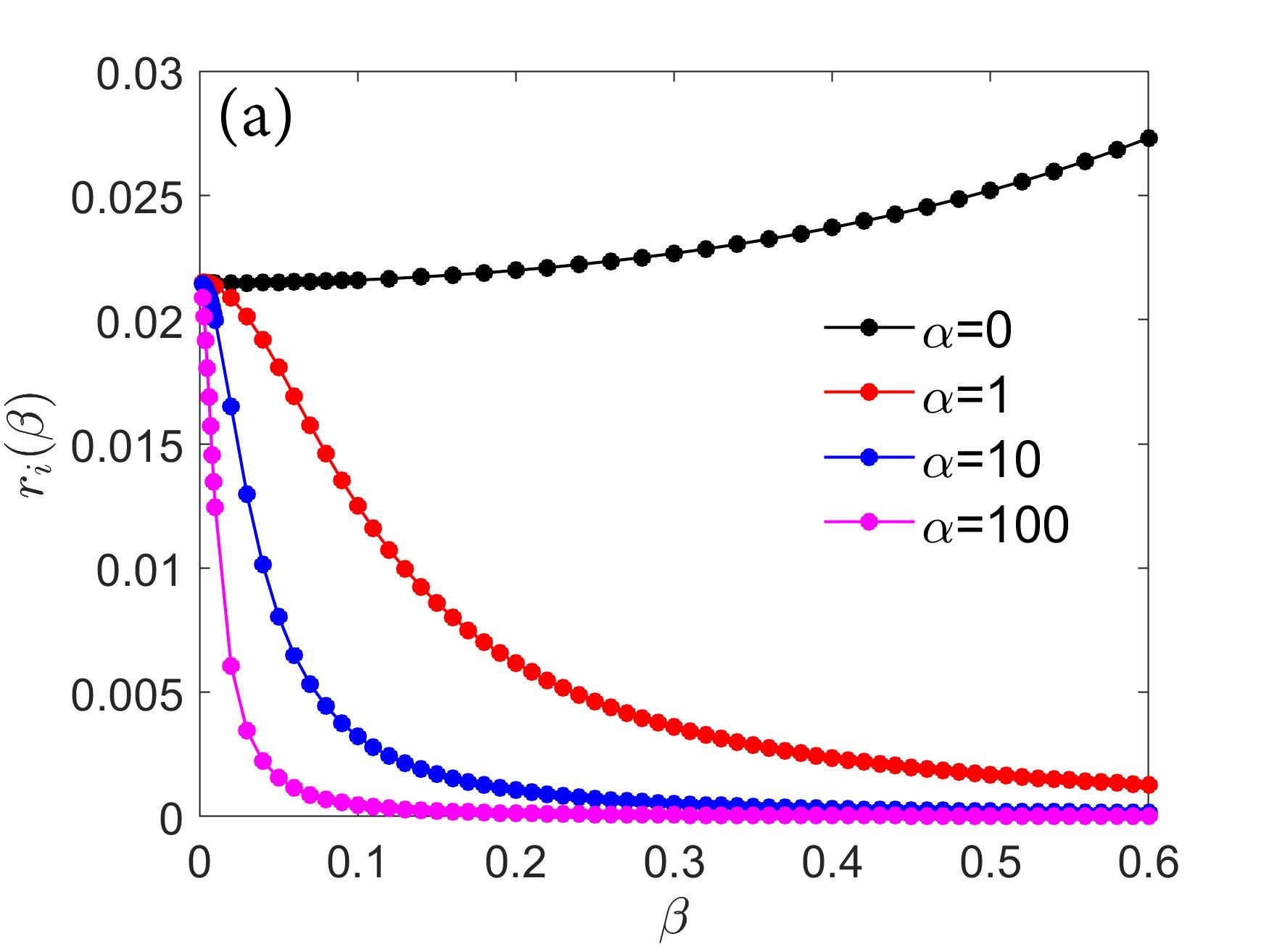}\\
	\includegraphics[width=\columnwidth]{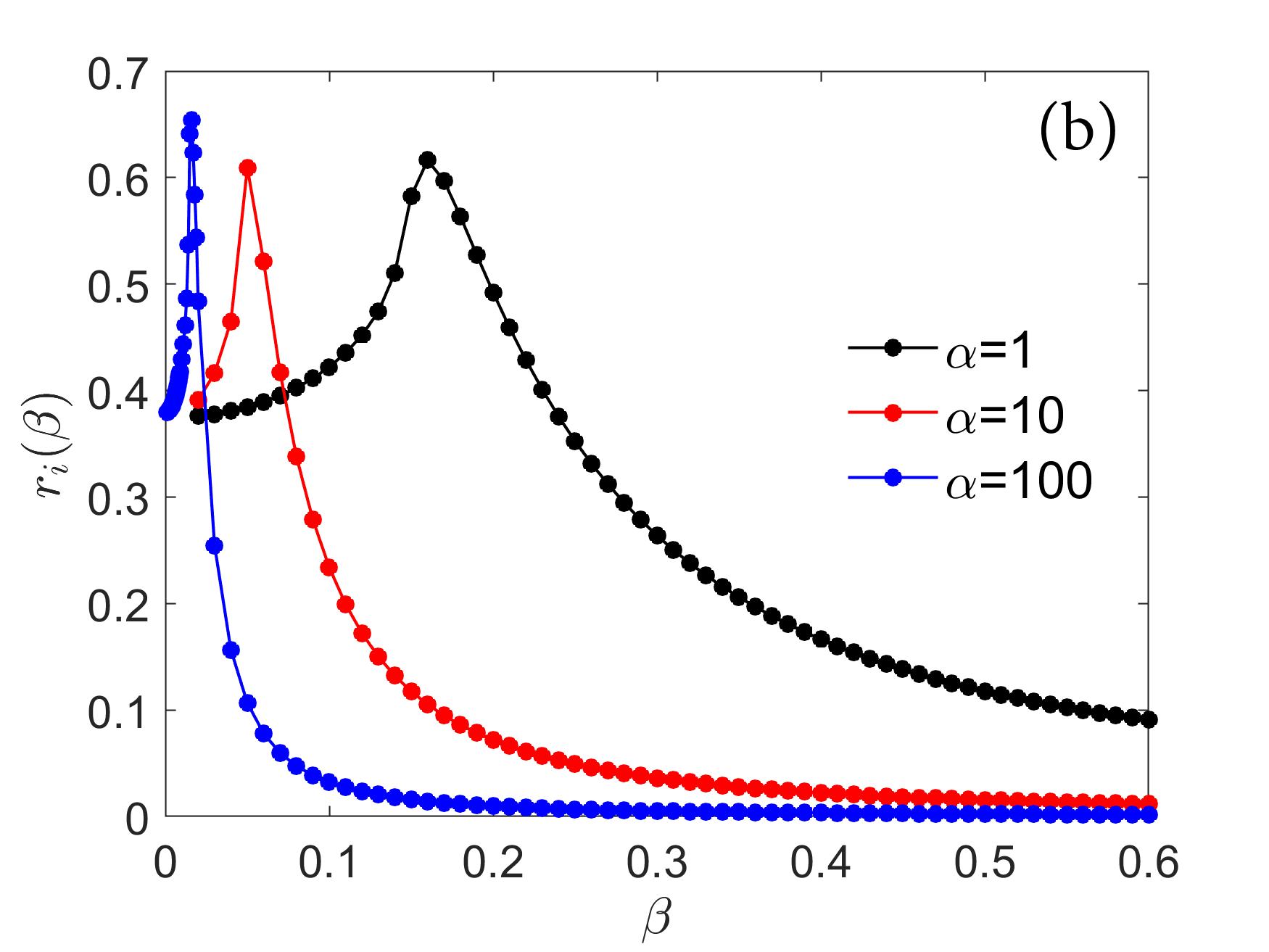}\\
	\includegraphics[width=\columnwidth]{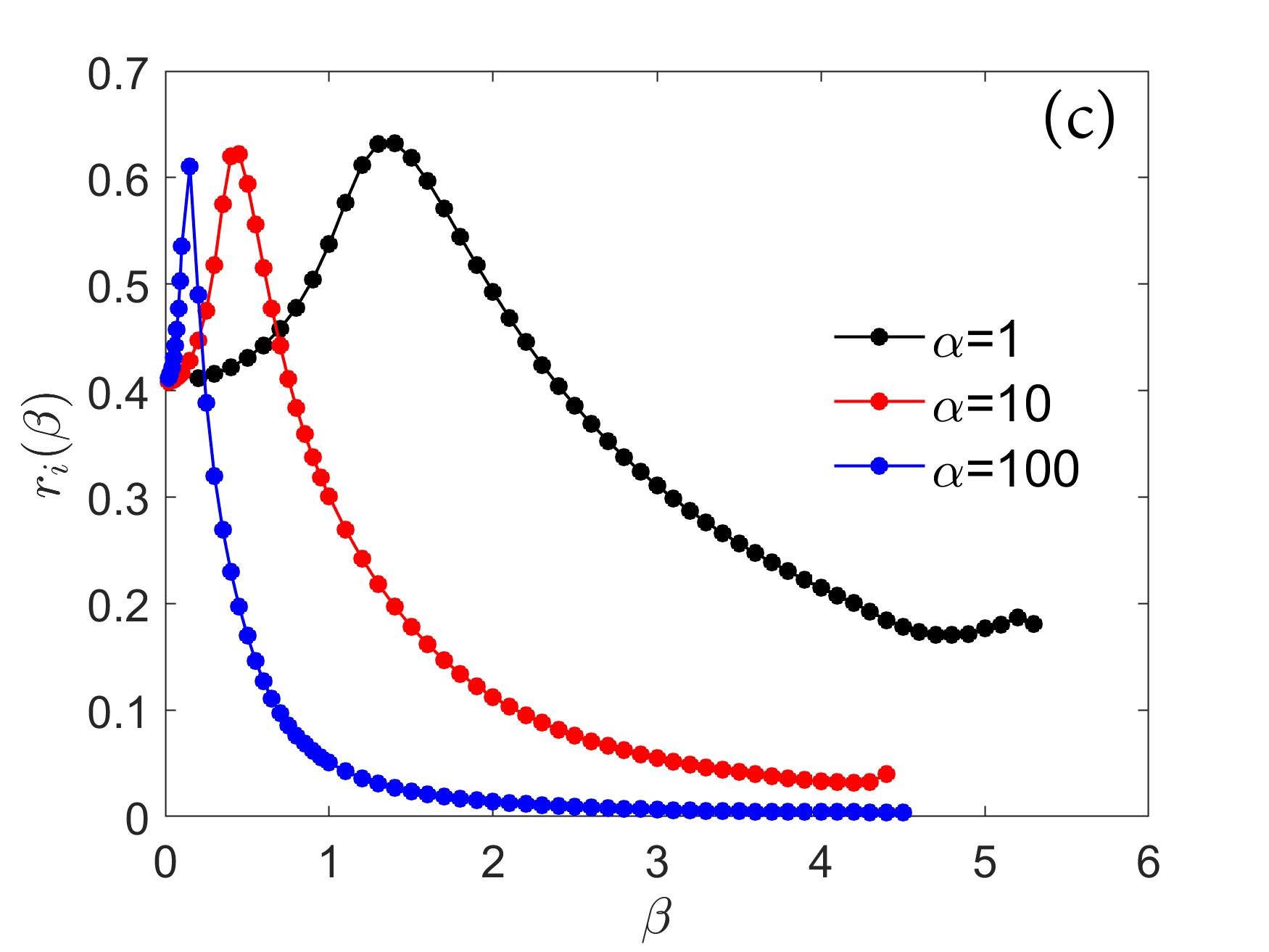}  
	\caption{$ r_i(\beta) $ calculated at $l/\lambda=3$ and (a) $\gamma=0.01$, $\alpha_0=10^4, 10^5, 10^6$, (b) $\gamma=0.1$, $\alpha_0=10^2,10^3, 10^4$, (c) $\gamma=1$, $\alpha_0=1,10,100$.}
	\label{fig:Fig4}
\end{figure}

Shown in Fig. \ref{fig:Fig4} are the field dependencies of $R_i(H)$ calculated at $l=3\lambda$ and different values of $\gamma$ and $\alpha_0$. At low frequency $\gamma=0.01$ and $\alpha_0=0$ the surface resistance increases with $\beta$ due to the effect of nonlinear vortex elasticity.  However, as $\alpha=\alpha_0\gamma^2$ increases, $r_i(\beta)$ starts decreasing with $\beta$ due to the decrease of the LO vortex viscosity with $v$, as it is evident from Eq. (\ref{p}).  Here $r_i(\beta)$ calculated numerically from Eq. (\ref{dyneq}) is in full agreement with the analytical results shown in Fig. \ref{fig:Fig2}.  

The behavior of $r_i(\beta)$ changes at higher frequencies, as shown in Figs.  \ref{fig:Fig4} (b) and (c) where $r_i(\beta)$ was calculated at $\gamma=0.1$ and $\gamma=1$. Here the field dependencies of $r_i(\beta)$ become nonmonotonic, the peaks in $r_i(\beta)$ shifting to lower fields as $\alpha_0$ increases.   At the peaks of $r_i(\beta)$, the velocity of vortex tip reaches the LO critical velocity $v_0$, but no jumps of the vortex tip occur because of the restoring effect of line tension of the vortex. Here the bending oscillations along the vortex are mostly confined within the elastic skin depth  $L_\omega$ given by Eq. (\ref{Lom}). If $\gamma=0.01$ the length $L_\omega \approx 4 \lambda $ is larger than $l=3 \lambda$, so the vortex segment swings as a whole and $r_i(\beta)$ decreases at all $\beta$. However, at $\gamma=0.1$ the length $L_\omega \approx 1.26\lambda$ is shorter than $l$ at $\beta\ll 1$. In this case $r_i(\beta) $ first increases with $\beta$, but after the peak in $r_i(\beta)$ at $\beta=\beta_p$ the velocity of the tip exceeds $v_0$ and the drag coefficient $\eta(v)$ drops rapidly with $v$, so $L_\omega\sim[\epsilon/\eta(v)\omega]^{1/2}$ becomes larger than $l$, and $r_i(\beta) $ starts decreasing with $\beta$ similar to the case shown in Fig. \ref{fig:Fig4} (a). At $\beta>\beta_p$ the nonlinear dynamics of the vortex becomes dependent on the vortex mass. Here the peaks in $ r_i(\beta) $ shift to higher $\beta$ as $\alpha = \alpha_0\gamma^2$ increases with frequency and the effect of the vortex mass  become more pronounced, so a stronger Lorentz forces is required to accelerate the vortex tip above the LO velocity. 

\begin{figure}
	\centering
	\includegraphics[width=\columnwidth]{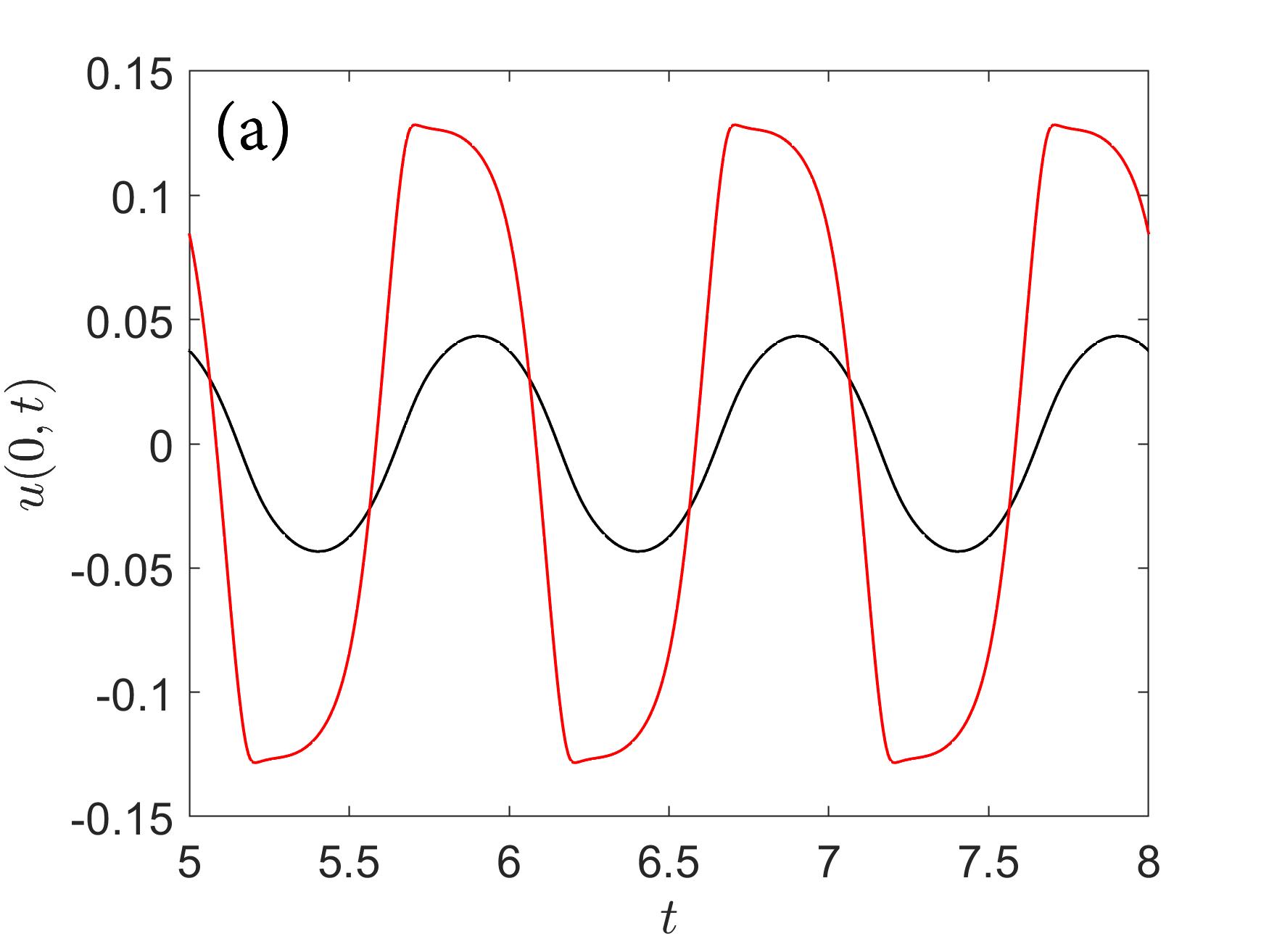}\\
	\includegraphics[width=\columnwidth]{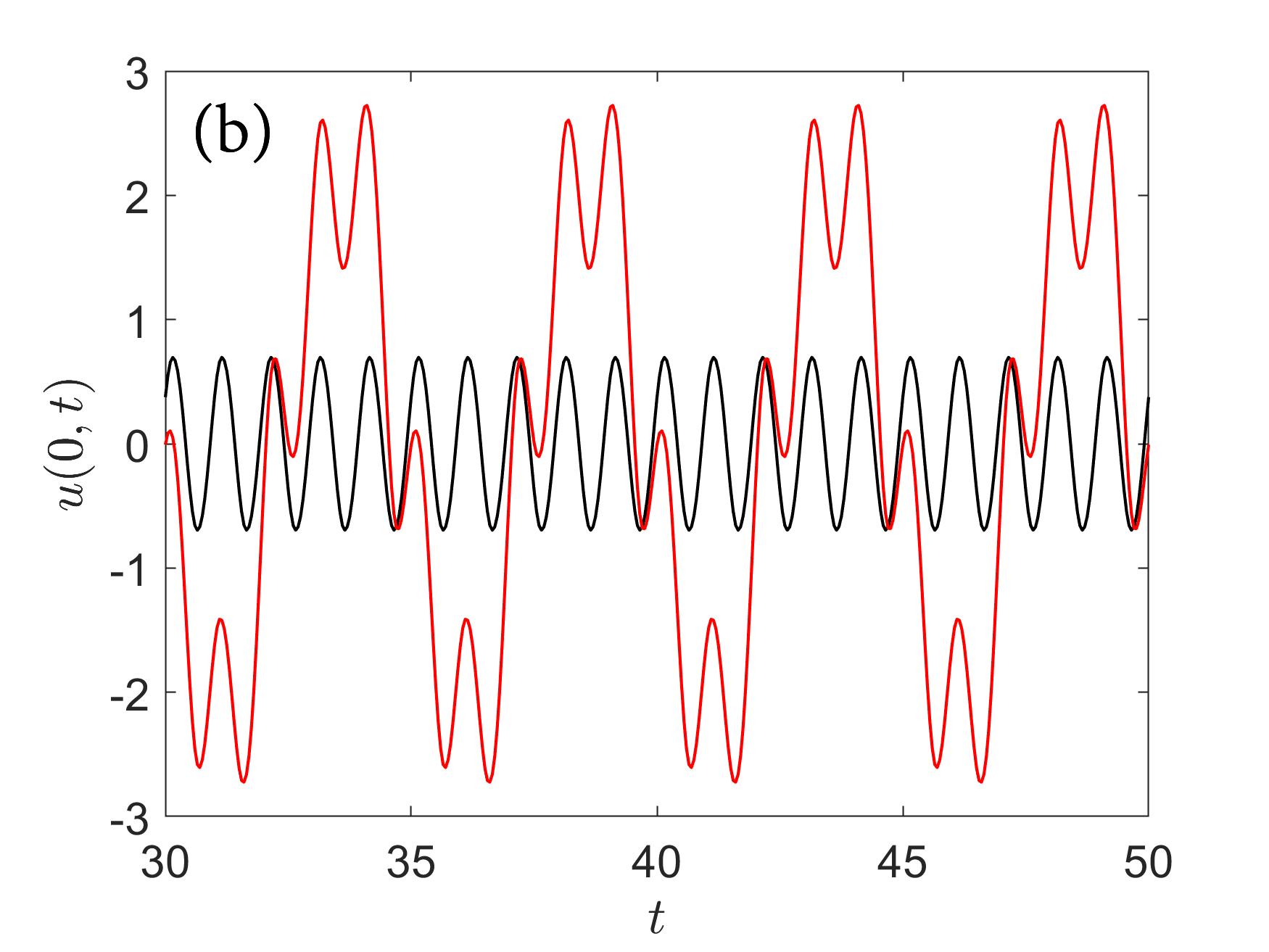} 
	\caption{Vortex tip oscillations:  
	(a) near the first peak in $r_i(\beta)$ shown in Fig. \ref{fig:Fig4}(b) at $\gamma=0.1$ and $\alpha_0=10^3$, (b) near the second peak in $r_i(\beta)$ shown in 
	Fig.  \ref{fig:Fig4}(c) at $\gamma=1$ and $\alpha_0=1$. Here the black and red lines correspond to $u(0, t) $ calculated at $\beta$ slightly below and above the peak.}
	\label{fig:Fig5}
\end{figure}

The dynamics of the vortex can change drastically once $\beta$ exceeds $\beta_p$. For instance, Fig. \ref{fig:Fig5}(a) shows the change in the time dependence of the vortex tip position $u(0,t)$ near the first peak in $r_i(\beta)$ at $\gamma=0.1$ and $\alpha_0=10^3$ in Fig. \ref{fig:Fig4}(b). Here  $u(0,t)$ changes from a nearly harmonic oscillations at $\beta<\beta_p$ to highly anharmonic oscillations at $\beta>\beta_p$ with a much greater amplitude of $u(0,t)$ due to a strong reduction of the local drag force at the tip at $v(0,t)>v_0$. The dynamics of $u(0,t)$ at $\beta>\beta_p$ resembles the van der Pol relaxation oscillations in a small mass limit ~\cite{bogolub}. As $\beta$ further increases, $u(0,t)$ becomes more harmonic because the effect of the LO nonlinear viscous drag diminishes. Yet a similar harmonic-anharmonic transition in $u(0,t)$ also happens above the second peak in $r_i(\beta)$, as shown in  Fig. \ref{fig:Fig5}(b). Here $u(0,t)$ was calculated at $\beta\simeq 5.2$,  $\gamma=1$, and  $\alpha_0=1$ corresponding to the second peak in $r_i(\beta)$ shown in Fig. \ref{fig:Fig4}(c).

\subsection{The effects of frequency and pin location}\label{fpin}

\begin{figure}[!htb]
	\centering
	\includegraphics[scale=0.12]{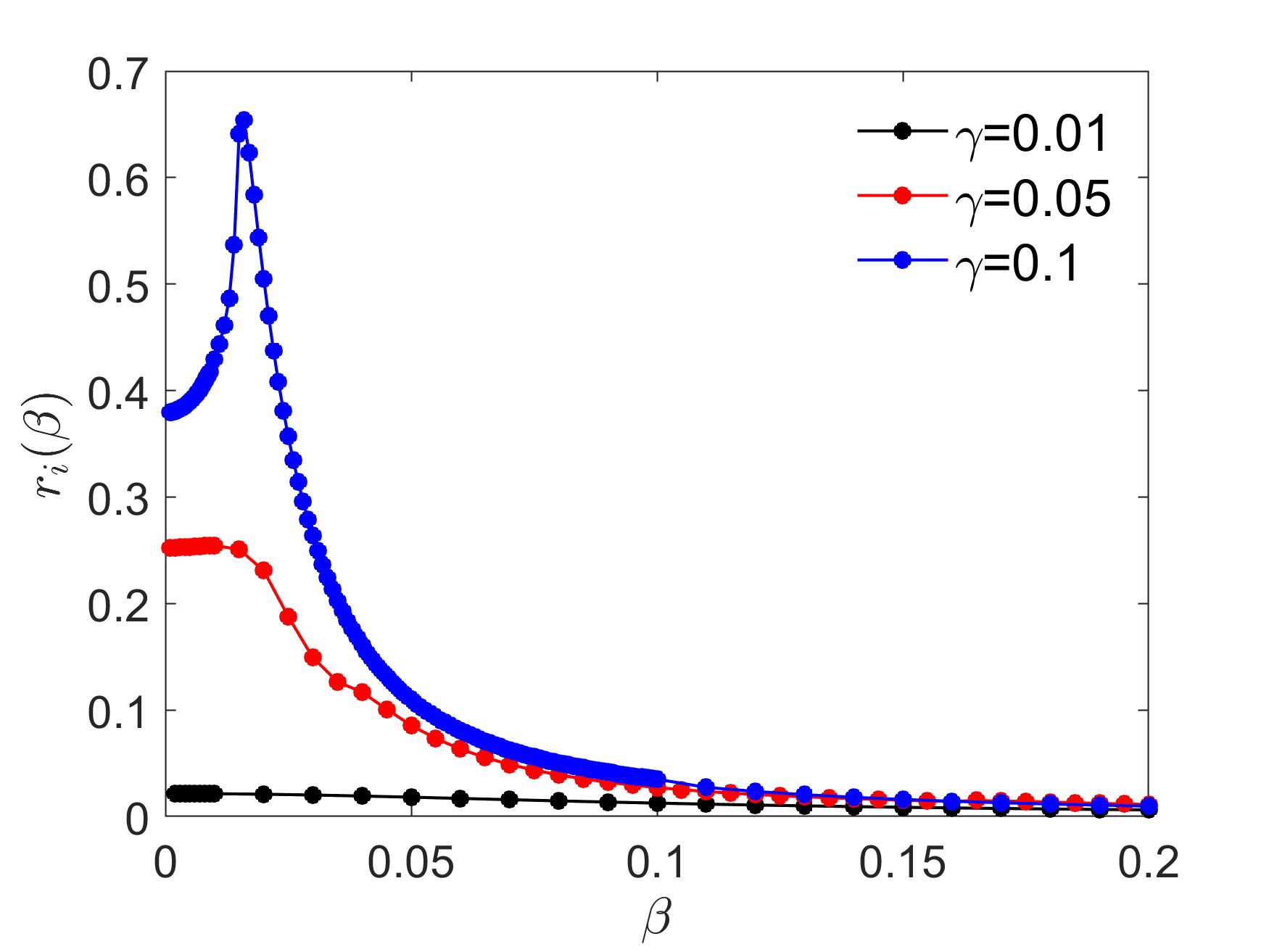}
	\caption{$ r_i(\beta) $ calculated  at $\gamma=0.01, 0.05, 0.1$, $\alpha_0=10^4$ and $l/\lambda=3$.}
	\label{fig:Fig6}
\end{figure}

The effect of frequency on the field dependence of $r_i(\beta)$ can be inferred from Fig. \ref{fig:Fig4}, using the frequency dependencies of the control parameters $\gamma \propto f$ and $\alpha \propto f^2$. For instance, Fig. \ref{fig:Fig6} shows the change in $r_i(\beta)$ at $\alpha_0=10^4$ as the  frequency increases. Here $r_i(\beta)$ is nearly field-independent at $\gamma=0.01$ but as the dimensionless frequency $\gamma$ increases, a strong decrease of $r_i(\beta)$ with the RF field develops.  This effect is a clear manifestation of the LO decrease of $\eta(v)$ with $v$, given that the velocity of the vortex increases as the frequency increases. Notice that the main drop in $R_i(H)$ occurs at small field amplitudes $H\lesssim 0.1H_{c1}$. 

\begin{figure}[!htb]
	\centering
	\includegraphics[scale=0.12]{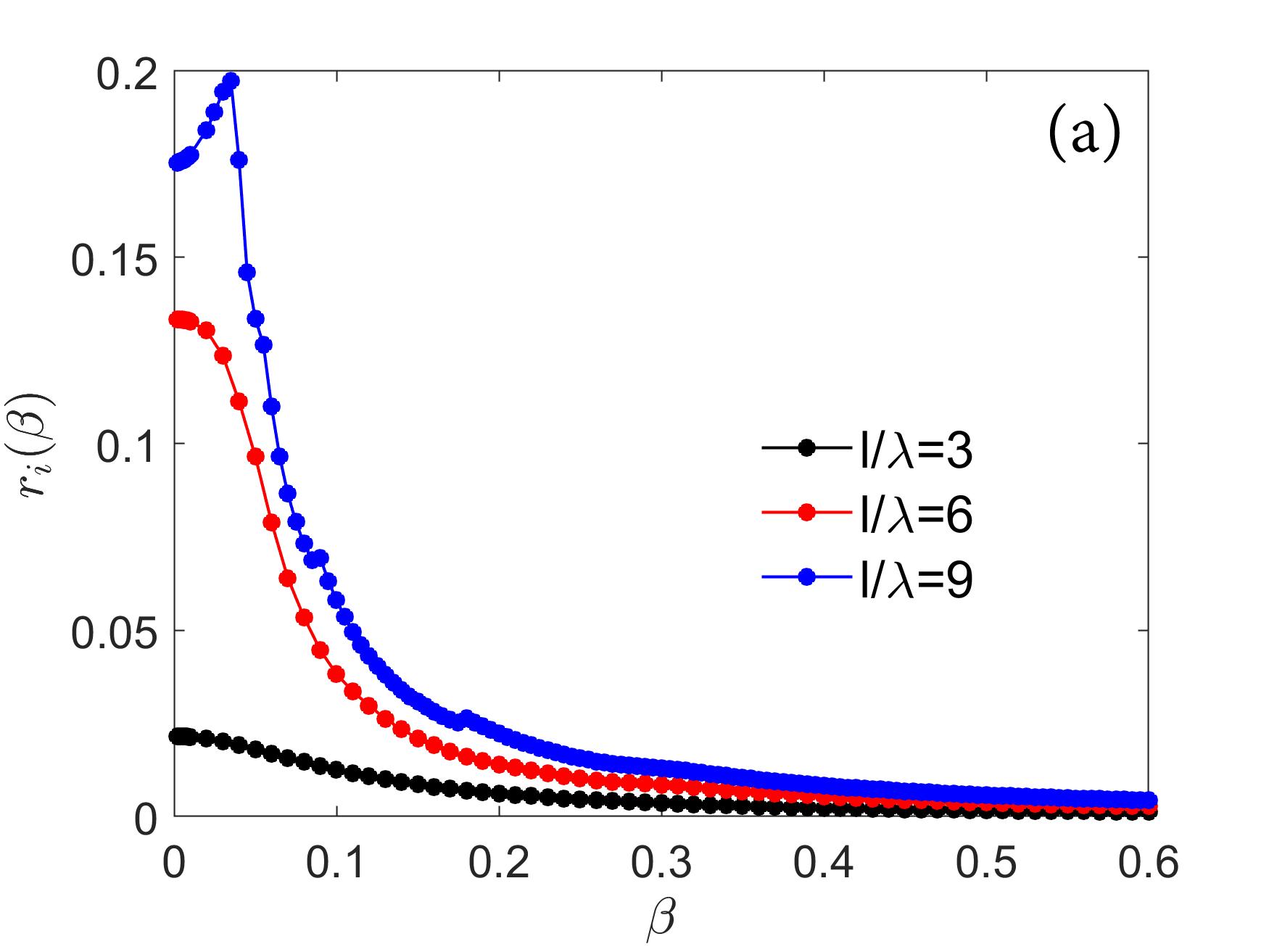}\\
	\includegraphics[scale=0.12]{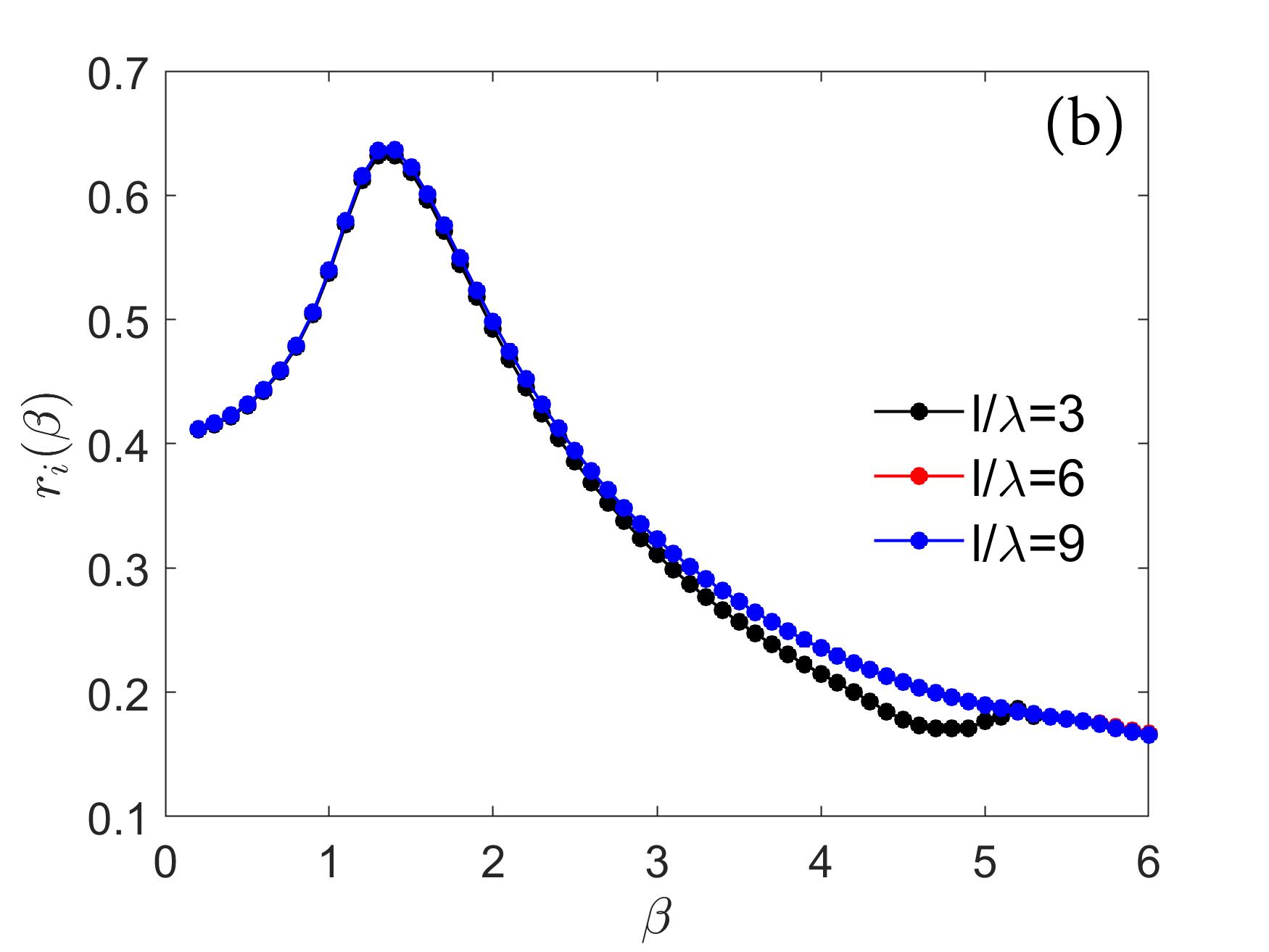}
	\caption{$ r_i(\beta) $ calculated at $\gamma=0.01$ (a), $\gamma=1$ (b) and different pin locations.}
	\label{fig:Fig7}
\end{figure}

Figures \ref{fig:Fig7} (a) and (b) show the effect of the pin position on the field dependence of $r_i(\beta)$ calculated at $\alpha=1$ and $\gamma=0.01$ and $\gamma=1$. As $\alpha$ increases the qualitative behavior of $r_i(\beta)$ remains the same but the peaks shift to smaller fields. Here $r_i(\beta) $ turns out to be sensitive to the pin locations at low fields and frequencies. This happens if $L_\omega$ at low $\beta$ and $\gamma$ exceeds $l$, so that the pinning center reduces the length of the oscillating vortex segment and thus the dissipation power. For example, at $l=9\lambda$ and $\gamma=0.01$, the low-field ripple length $L_\omega=4\lambda$  is shorter than $l$, and the non-monotonic behavior of $r_i(\beta)$ is similar to that is shown in Fig. \ref{fig:Fig4}. As $\beta$ increases $\eta(v)$ decreases and the nonlinear ripple length $L_\omega$ becomes much larger than $l$. In this case the vortex segment of length $l$ swings as a whole, and $r_i$ given by Eq. (\ref{univ}) is proportional to $l$ and decreases with $\beta$. At high frequency, $\gamma=1$, the low-field ripple length $L_\omega$ is shorter than $l=(3-9)\lambda$ used in our simulations, and $r_i(\beta) $ becomes practically independent of the pin location except for a small second hump in $r_i(\beta)$ at $\beta\simeq 4.7$  for $l=3\lambda$.

\subsection{The effect of the mean free path.}
  
The nonlinear dynamics of the trapped vortex and the field dependence $r_i(\beta)$ can be tuned by nonmagnetic impurities because the control parameters $\gamma$, $\alpha$, $\beta$ and $\mu$ defined by Eqs. (\ref{gamma})-(\ref{mu}) increase strongly as the mean free path $l_i$ decreases.  Using $\rho_n\propto l_i^{-1}$, $ v_0 \propto l_i^{1/2}$ ~\cite{LO} and the conventional GL interpolation formulas $\lambda=\lambda_0\Gamma$,  and $\xi=\xi_0/\Gamma$, where $\Gamma=(1+\xi_0/l_i)^{1/2}$, the explicit dependencies of $\alpha$ $\gamma$ and $\beta$ on $l_i$ can be presented in the form:
\begin{gather}
\gamma=\frac{g_0\Gamma^6 l_if}{g\xi_0f_0}, \qquad \alpha =\frac{\xi_0}{l_i}\left(\frac{\lambda_0 f}{\tilde{v}_0}\right)^2,\\
\beta=\frac{ g_0\Gamma^2H}{gH_{c10}},\qquad 
\mu =\frac{\lambda_0^2f^2 M\Gamma^4g_0}{\phi_0 gH_{c10}},\\
g_0=\ln\frac{\lambda_0}{\xi_0}+\frac{1}{2},\qquad g=\ln\frac{\lambda_0\Gamma^2}{\xi_0}+\frac{1}{2}.
\end{gather}
Here $f_0$ is defined by Eq. (\ref{gamma}), where $\lambda_0$, $\xi_0$ and $H_{c10}=\phi_0g_0/4\pi\mu_0\lambda_0^2$ are the penetration depth, coherence length and the lower critical field in the clean limit, respectively, and $\tilde{v}_0$ is the LO critical velocity at $l_i=\xi_0$.  For the sake of simplicity, we assume that $M$ is independent of $l_i$, and the Bardeen-Stephen formula for $\eta_0$ can be used in a moderately clean limit as well ~\cite{kopnin}.

Equation (\ref{dyneq}) was solved for dirty Nb using $\lambda_0=\xi_0=40$ nm, $\tilde{v}_0=126$ m/s~ \cite{grimaldi2008}, $\gamma_0=g_0f/f_0\approx 0.004$ and $\lambda_0^2f^2/\tilde{v}_0^2\approx 0.1 $ at 1 GHz. We calculated $r_i(\beta)$ for different values of the mean free path $l_i/\xi_0=1,0.1$ and $0.05$ at frequencies 1 GHz, 10 GHz and 100 GHz, and $\mu/\gamma=8 \cdot 10^{-4}(\xi_0/(\Gamma^2l_i))$ at 1 GHz. As $l_i$ decreases the parameter $\alpha\propto l_i^{-1}$ increases and the surface resistance starts decreasing with $H$, the main drop of $r_i(H)$ shifting to lower fields as the material gets dirtier. 
\begin{figure}[!htb]
	\centering
		\includegraphics[width=\columnwidth]{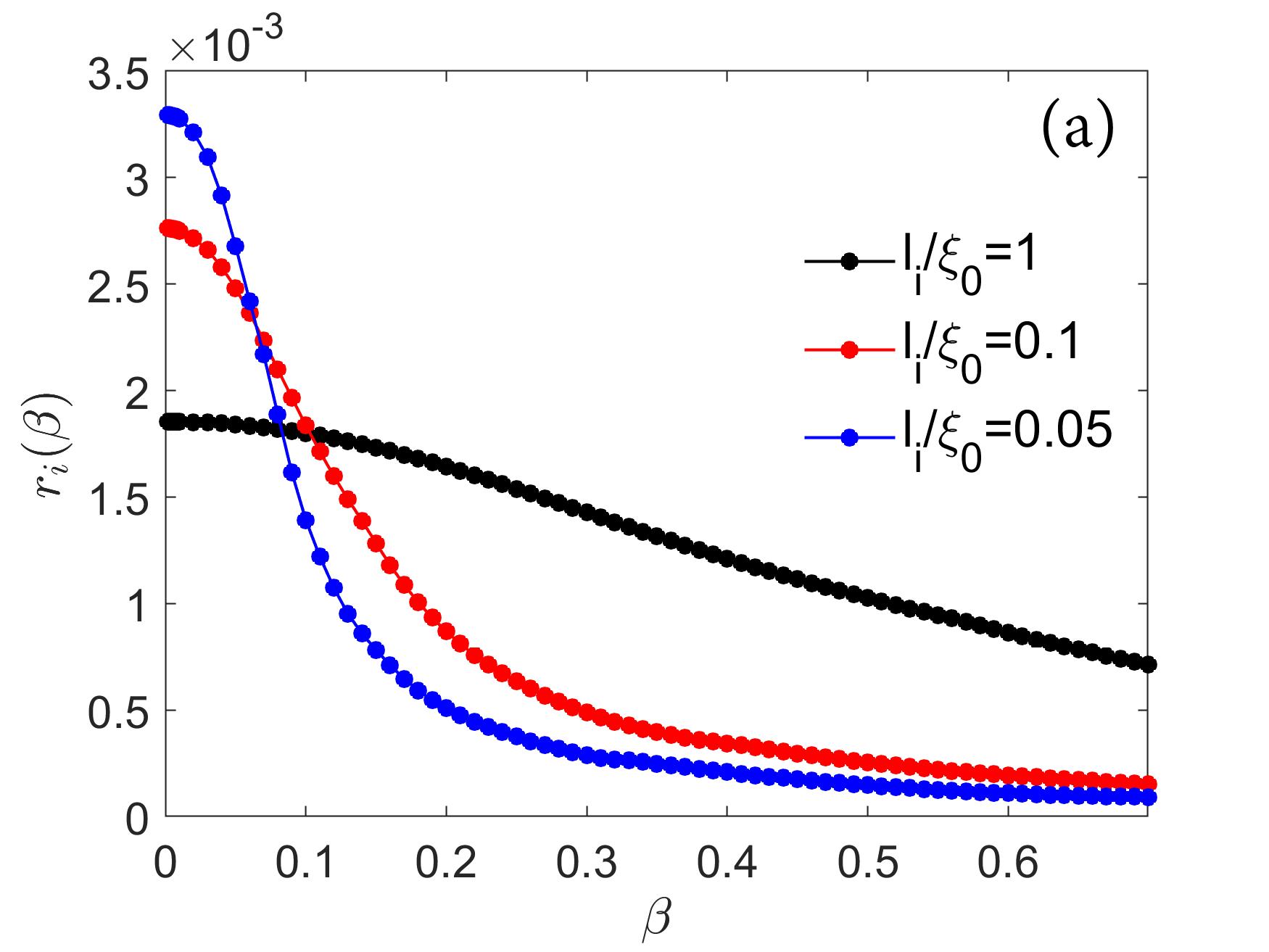}\\
		\includegraphics[width=\columnwidth]{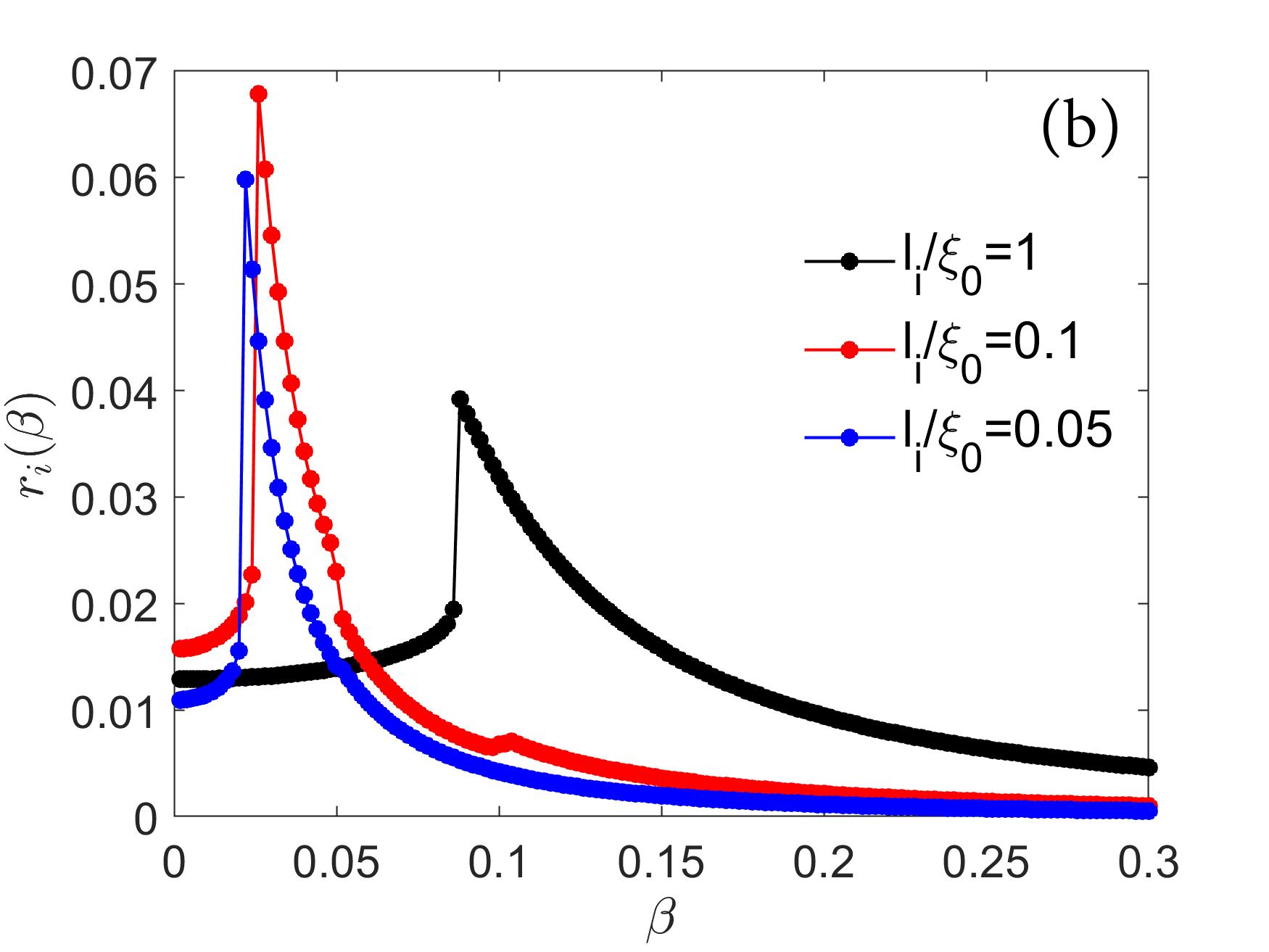}\\
		\includegraphics[width=\columnwidth]{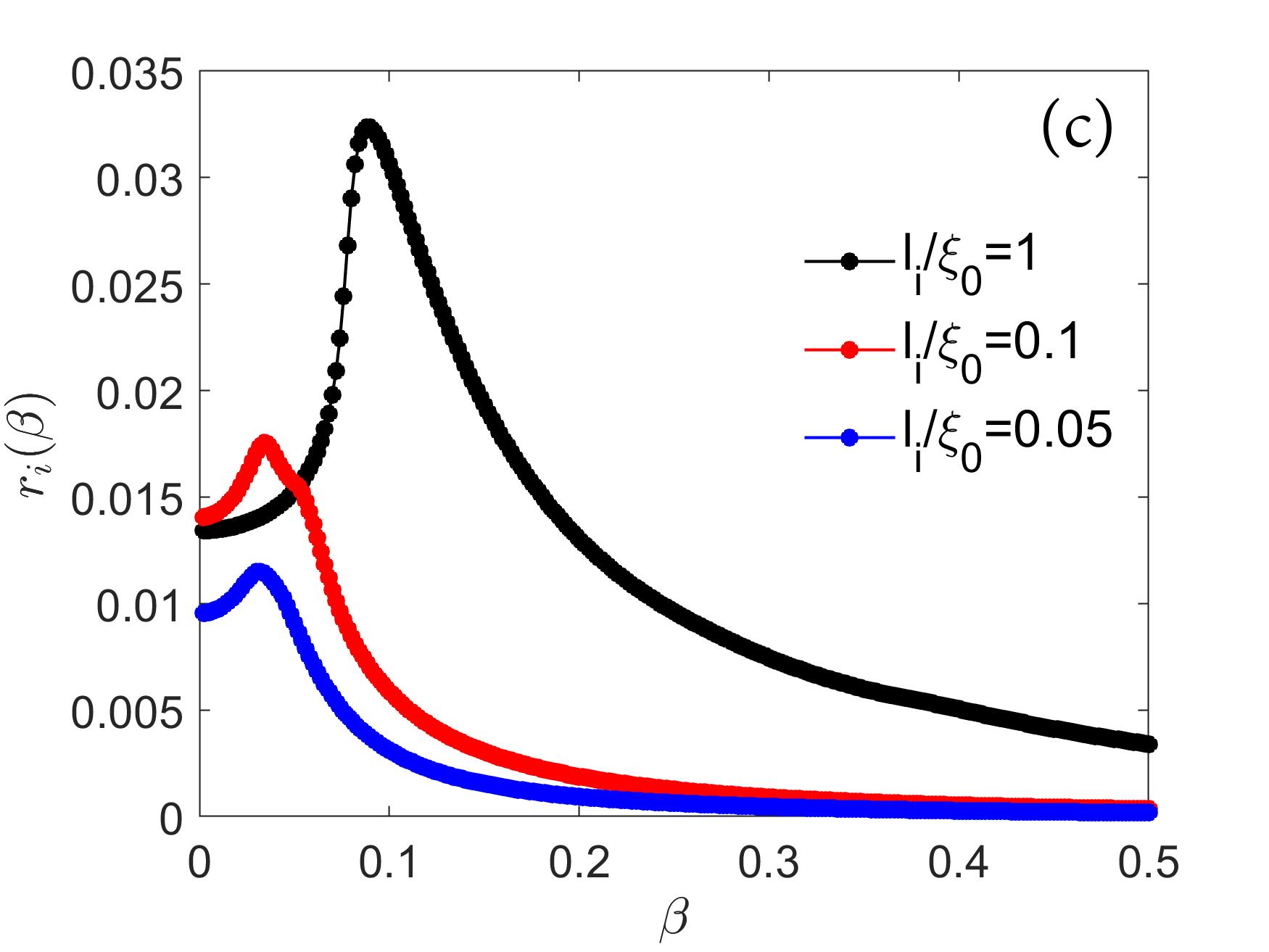}  
	\caption{The surface resistance $r_i(\beta) $ calculated at: (a) $ \gamma_0=0.004$, $\alpha_0=0.1, $(b) $\gamma_0=0.04$, $\alpha_0=10$, (c) $ \gamma_0=0.4$, $\alpha_0=1000$ for different values of $l_i/\xi_0$ and $l/\lambda_0=3$.}
	\label{fig:Fig8}
\end{figure}

Figure \ref{fig:Fig8} shows $r_i(\beta)$ calculated for different values of frequency $\gamma$ and the mean free path $l_i$. At the lowest frequency $\gamma_0=0.004$ the curves $r_i(\beta)$  shown in Fig. \ref{fig:Fig8} (a) exhibit the monotonic decrease with $\beta$ similar to that was discussed in Sec. \ref{lf} in the low-$\gamma$ limit of the ripple length $L_\omega$ exceeding the pin distance $l$. As the frequency increases a nonmonotonic field dependence of $r_i(\beta)$ develops, the peaks in $r_i(\beta)$ shifting to lower fields as the ratio $l_i/\xi_0$ decreases. Notice the jumps preceding the peaks in $r_i(\beta)$ in the case of $\gamma_0=0.04$ shown in Fig. \ref{fig:Fig8} (b).  The nonmonotonic dependence of $r_i(\beta)$ is a manifestation of the transition from the case of $L_\omega <l$ at low fields to $L_\omega>l$ at higher fields, as was discussed above.  In the case of $\gamma_0=0.4$ represented by Fig. \ref{fig:Fig8} (c) the nonmonotonic field dependences of $r_i(\beta)$ remain qualitatively similar to those shown in Fig. \ref{fig:Fig8}(b), except that the peaks in $r_i(\beta)$ get broadened and the jumps in $r_i(\beta)$ characteristic of $\gamma_0=0.04$ disappear.  The latter results from the effect of the vortex mass, since the contribution of the inertial term in Eq. (\ref{dyneq}) becomes more pronounced at higher frequencies. Generally, the effect of mass smoothes sharp jumps characteristic of nonlinear relaxation oscillations ~\cite{bogolub}.  Overall, the evolution of $r_i(\beta)$ with $\gamma$ at different mean free paths shown in Fig. \ref{fig:Fig8} (b) and (c) appears similar to that of $r_i(\beta)$ calculated for different pin spacings (see Fig.  \ref{fig:Fig7}).  

\begin{figure}[!htb]
	\centering
		\includegraphics[width=\columnwidth]{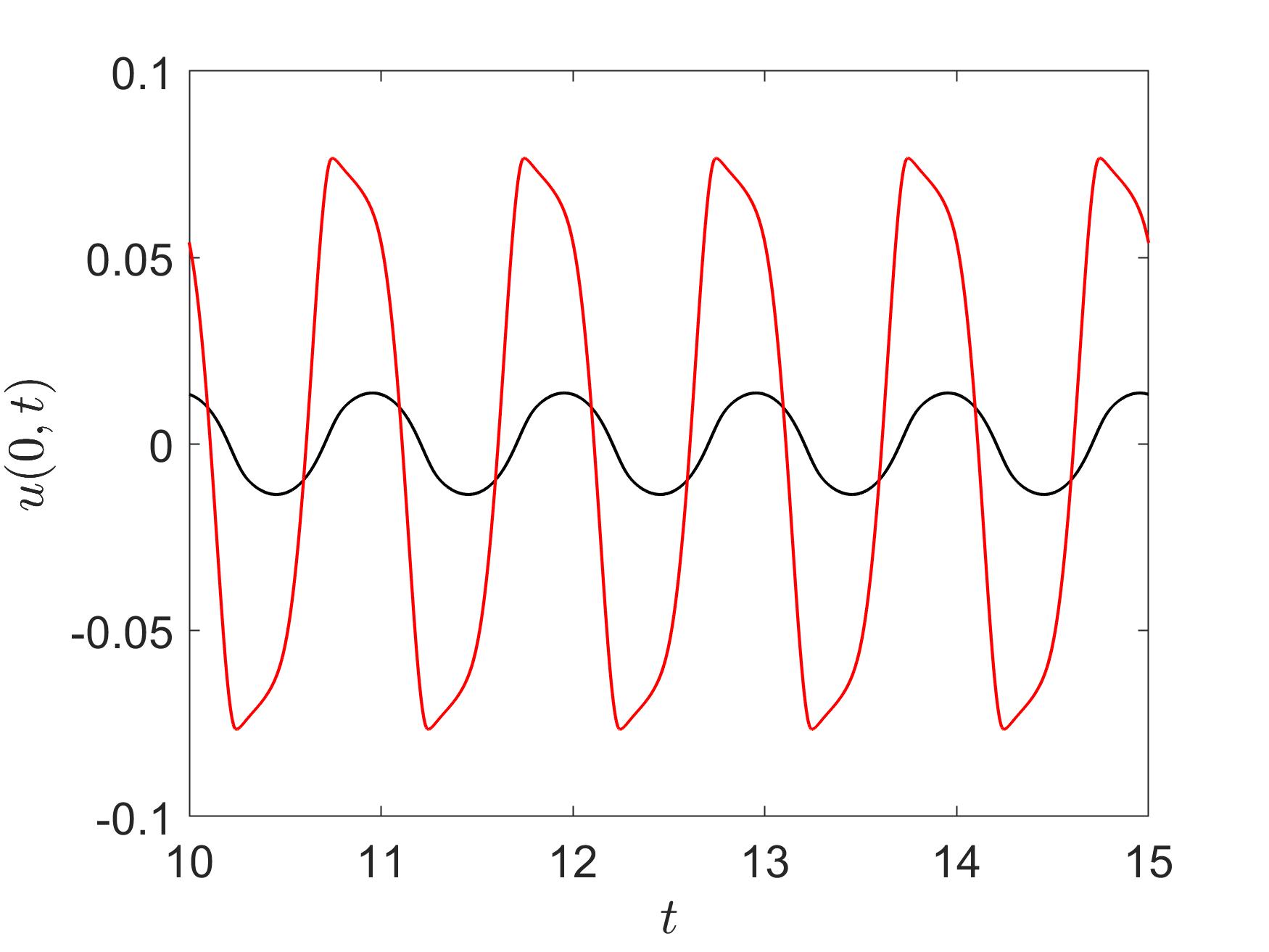}  
	\caption{Vortex tip oscillations at the jump in $r_i(\beta) $ shown in Fig. \ref{fig:Fig8} (b) for the case of $l=3\lambda_0$, $\gamma_0=0.04$, 
	$ l_i/\xi_0 = 0.1 $ and $\alpha=100$. Here the black and the red lines correspond to $u(0, t) $ at $\beta<\beta_p$ and $\beta>\beta_p$. }
	\label{fig:Fig9}
\end{figure}

Figure \ref{fig:Fig9} shows how the time dependence of the vortex tip position $u(0, t)$ changes from nearly harmonic oscillations at $ \beta<\beta_p$ to relaxation oscillations at $\beta>\beta_p$. This dynamic transition occurs at small fields, for example, $\beta>\beta_p\approx 0.086$ at $f = 10$ GHz and $ l_i/\xi_0 = 1 $. At higher frequencies the change in $u(0,t)$ near the rounded peaks in $r_i(\beta)$ shown in Fig. \ref{fig:Fig8} (c) becomes less pronounced, turning into a gradual increase of anharmonicity in $u(0,t)$ as $\beta$ is increases from $\beta<\beta_p$ to $\beta>\beta_p$.  

\section{Dynamic kinks along a vortex.}

The vortex line tension suppresses the jumpwise LO instability which nevertheless manifests itself in the anomalous decrease of the surface resistance $r_i(\beta)$ with the rf field amplitude at $\beta>\beta_p$. Yet a principal question remains whether there is a range of the parameters in which the nonmonotonic LO velocity dependence of the drag force could cause a dynamic shape instability of a moving vortex.  Indeed, once the velocity of a small vortex segment exceeds $v_0$, the local vortex drag diminishes further increasing $v(x,t)$ and resulting in growing shear stress between the fast vortex tip at $x\lesssim \lambda$ and a slower part of the vortex at $x\gtrsim\lambda$. In this section we show that at large enough frequencies dynamic solutions of Eq. (\ref{dyneq}) become singular, formally indicating a vortex teardown as the restoring effect of vortex line tension cannot counter the dynamic LO shear stress at strong rf drives $\beta>\beta_c(\gamma,l)$.

\begin{figure}[!htb]
	\centering
	\includegraphics[scale=0.3]{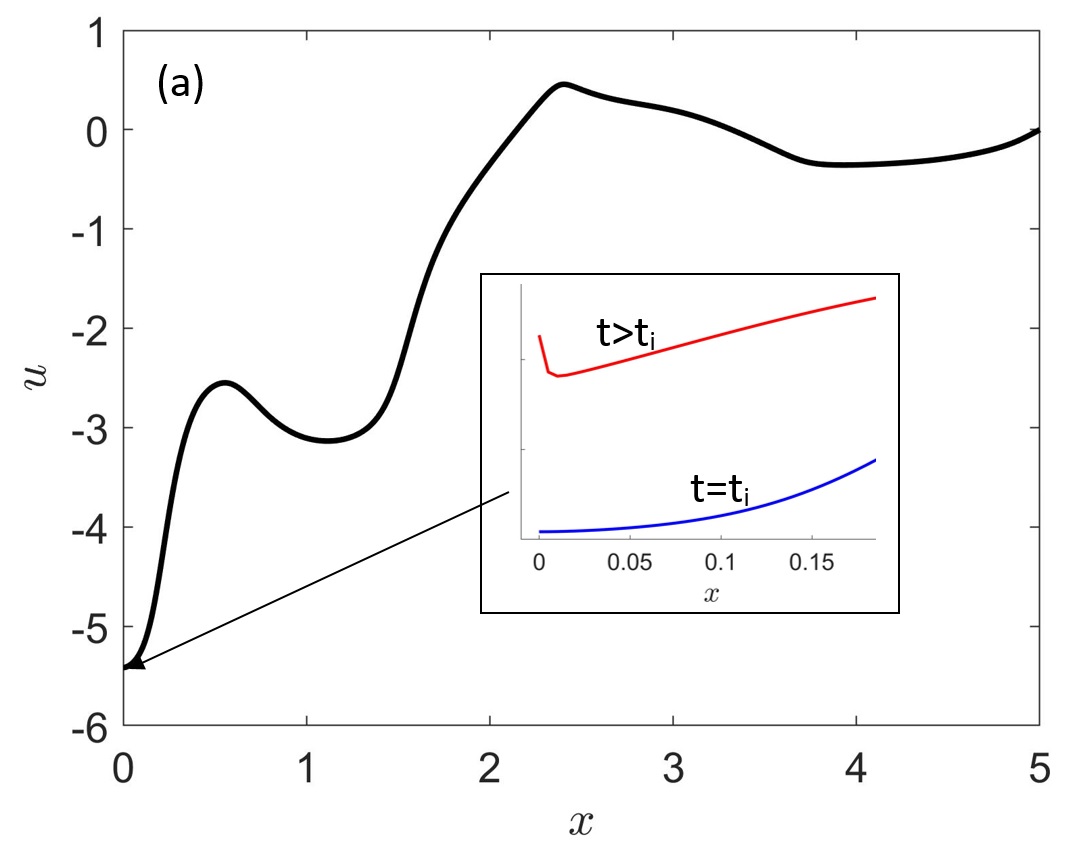}\\
	\includegraphics[scale=0.3]{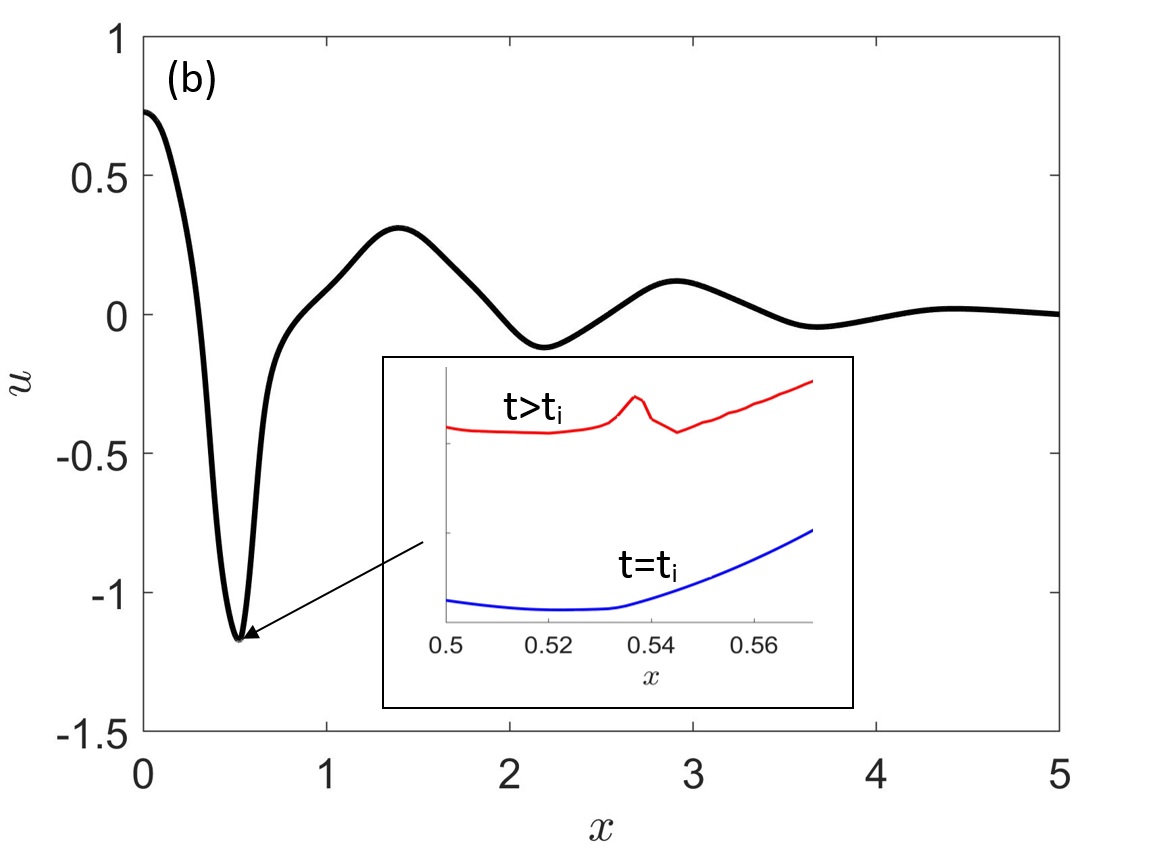}
	\caption{Dynamics of the vortex teardown calculated at $l=5\lambda$ and: (a) $\gamma=1.2$ and $\alpha_0=1$ for which $\beta_c = 9.8$, (b) $\gamma=2.4$ and $\alpha_0=1$ for which  	$\beta_c=15.8$. Insets show the dynamics of the vortex filament at the teardown point $x=l_c$ after the instant of the instability $t_i$ (the curves $u(x,t)$ at $t=t_i$ and $t>t_i$ are shifted for clarity).}
	\label{fig:Fig10}
\end{figure}

Shown in Fig. \ref{fig:Fig10} are examples of the vortex shape instability which can happen both at the surface and at the point $x=l_c<l$ between the surface and the pin position at $l=5\lambda$. The instability first develops as a cusp at $x=l_c$ which then evolves into a growing jump in $u(x,t)$, as shown in the insets. The resulting large derivative $u'(x,t)$ at $x=l_c$ reduces the restoring effect of the line tension in Eq. (\ref{dyneq}), further facilitating the vortex teardown. Because of large curvature of $u(x,t)$ at $x=l_c$, the elastic response becomes nonlocal and the assumption that the line tension $\epsilon$ in Eq. (\ref{dyneq}) is independent of the wave  vector $k$ of bending distortion along the vortex fails.  In a linear elasticity theory $\epsilon(k)$ in a uniaxial superconductor becomes dependent on $k$ at $k\lambda\gtrsim 1$ ~\cite{blatter,ehb}:  
\begin{equation}
\epsilon (k)=\frac{\epsilon _{0}}{2\Xi}\ln \frac{\kappa ^2\Xi}{1+\lambda
^{2}k^{2}}+\frac{\epsilon _{0}}{2\lambda ^{2}k^{2}}\ln (1+\lambda^{2}k^{2}),
\label{eps}
\end{equation}
where $\epsilon _{0}=\phi _{0}^{2}/4\pi \mu _{0}\lambda ^{2}$ and $\Xi=\lambda_c^2/\lambda^2$ is the band electron mass anisotropy parameter. At $k\lambda\ll 1$ and $\Xi=1$ Eq. (\ref{eps}) yields $\epsilon=\epsilon_0 g$ used in our simulations, but at $k\lambda\gg 1$, the line tension $\epsilon(k) \simeq -\epsilon_0\ln(\xi k)$ decreases slowly as $k$ increases up to $k\simeq \xi^{-1}$.  Yet because the vortex becomes softer for short wavelength distortions with $k\lambda\gtrsim 1$, the elastic nonlocality may  facilitate the shape instability at the cusp at $x=l_c$ where $k\lambda\gg 1$ and decrease $\beta_c$ as compared to $\beta_c$ calculated here at $k\lambda\ll 1$.  It turned out that the development of the shape instability is also affected by the vortex mass.  

\begin{figure}[!htb]
	\centering
	\includegraphics[scale=0.12]{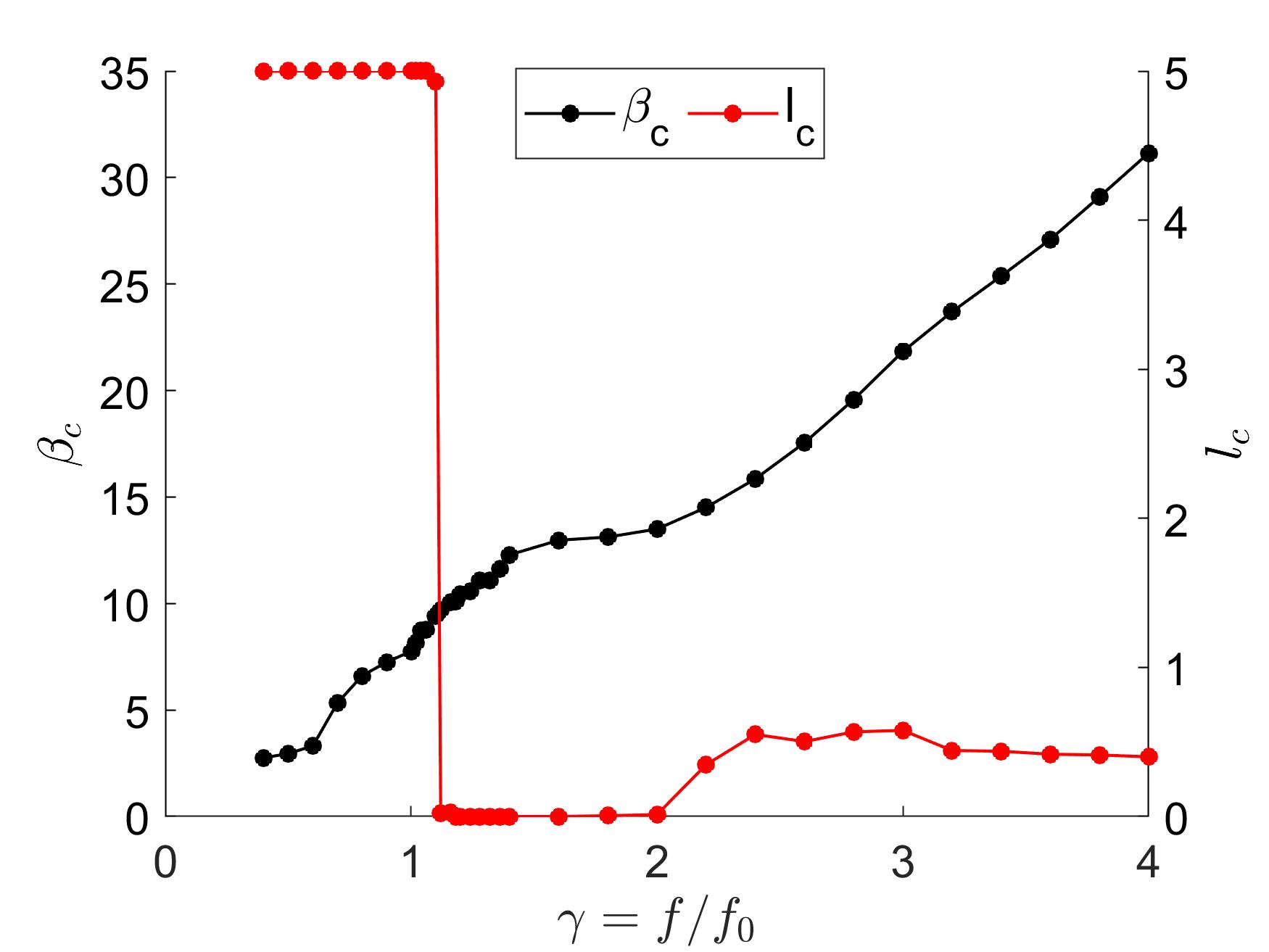}
	\caption{Frequency dependencies of $l_c$ and $\beta_c$ calculated at $\alpha_0=1$, $\mu_1=0.08$, and $l=5\lambda$. }
	\label{fig:Fig11}
\end{figure}

Shown in Fig. \ref{fig:Fig11} is an example of the frequency dependencies of the instability coordinate $l_c$ and the critical field $\beta_c$ calculated at $\alpha_0=1$ and $\mu_1=0.08$. In these simulations each value of $\beta_c(\gamma)$ was calculated by slowly ramping up the field $\beta(t)=0.01t$ at $0.4<\gamma<1.2$ and $\beta(t)=0.05t$ at $1.2<\gamma<4$, and defining $l_c$ at a point where $u'(l_c,t)$ reaches  $u'_c=100$.  The results show that $\beta_c (\gamma)$ increases monotonically with $\gamma$ while $l_c(\gamma)$ exhibits a non-monotonic dependence with a jump at $\gamma \approx 1.1$. 

The behavior of $l_c(\gamma)$ can be understood as follows. At low $\gamma$ the critical gradient $u'(l_c,t)>u_c$ first occurs at the pin position, so the vortex instability is just the quasi-static depinning considered in Sec. \ref{lf}. As $\gamma$ increases, the ripple length $L_\omega(\gamma,\beta)$ becomes  smaller than $l$ so the effect of the pin on the dynamics of the vortex weakens. As a result, the mechanism of the vortex shape instability changes from the quasi-static depinning to the dynamic LO shear instability at the surface, $l_c=0$, where the Meissner current density is maximum. At moderate frequencies $1\lesssim \gamma\lesssim 2$ the vortex tip has the highest velocity $v(0,t)$ at the surface, where the strongest shear gradient develops if $v(0,t)$ becomes much larger than $v_0$.  

As $\gamma$ further increases, the instability point moves from the surface to a finite $l_c\simeq 0.5\ll l$, as shown in Fig. \ref{fig:Fig11}. The shape of the vibrating vortex line evolves from a monotonic $u(x)$ at $\gamma\ll 1$ to an oscillatory $u(x)$ at $\gamma\gtrsim 1$, as illustrated by Fig. \ref{fig:Fig10}. Because of spatial oscillations in $u(x,t)$, the maximum shear gradient in $u(x,t)$ at large $\gamma$ occurs in the bulk rather than at the surface. Since the instability point $l_c\simeq 0.5$ is far away from the pin position $l=5$, the shape instability is not affected by pinning but is mostly controlled by the nonlinear LO viscosity, vortex elasticity and the mass $M$. Our simulations have shown that the shape instability is affected by the ramp rate of $\beta(t)$, but its effect is rather mild and does not change the qualitative behaviors of $\beta_c(\gamma)$ and $l_c(\gamma)$. 

\begin{figure}[!htb]
	\centering
	\includegraphics[width=8cm,trim={120mm 35mm 120mm 35mm},clip]{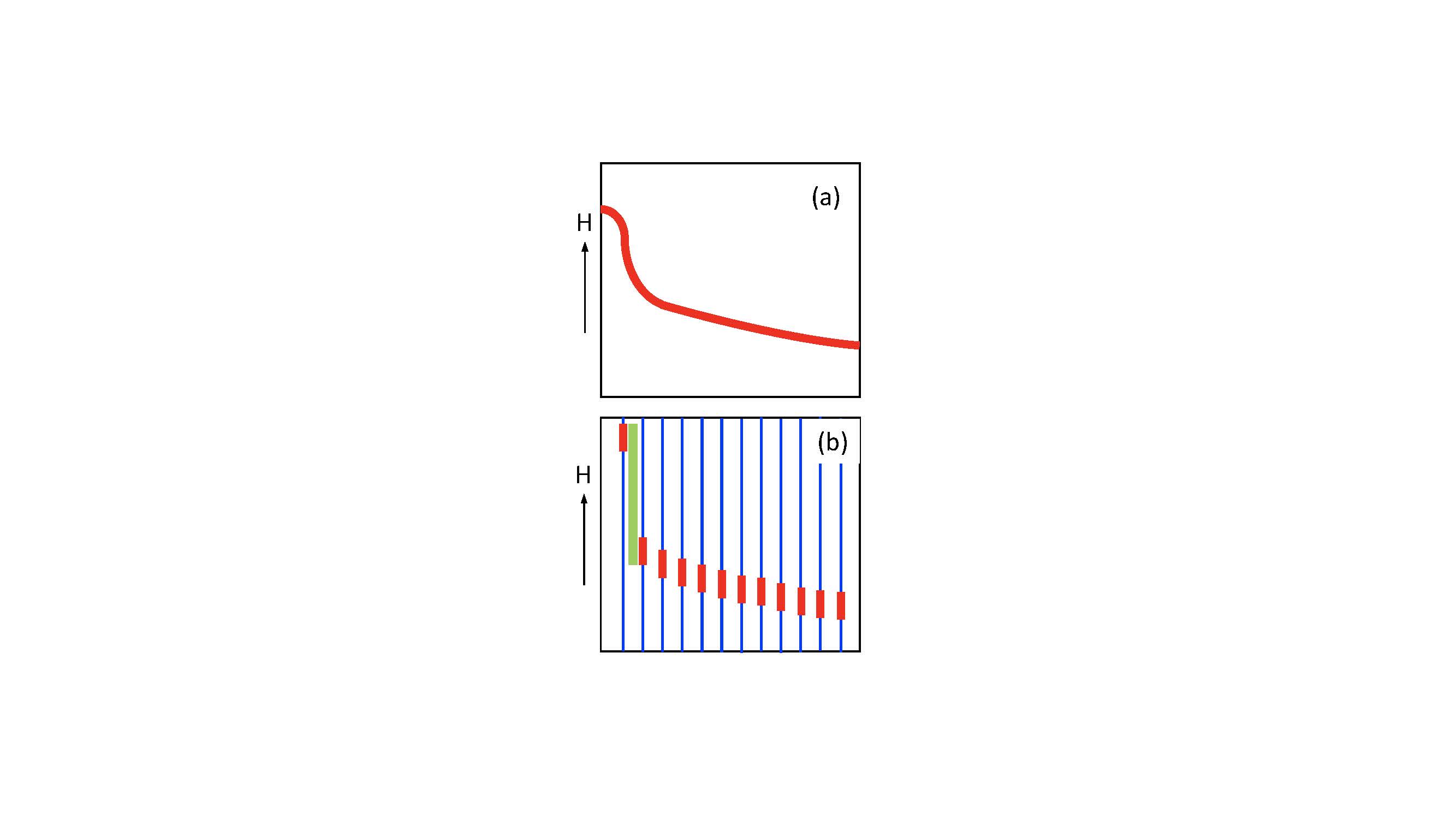}
	\caption{Snapshots of dynamic kinks along a curvilinear vortex driven by a strong ac magnetic field: (a) A kink along the continuous vortex core (red) at the surface of an isotropic superconductor.  (b) A vortex composed of a stack of pancake vortices in a layered superconductor. The green rectangle shows a Josephson string between the runaway pancake vortex with $v>v_0$ at the surface and a slower 2D vortex stack.  }
	\label{fig:Fig12}
\end{figure}

The above singular solutions $u(z,t)$ describe the formation of a cusp on a vortex which then develops into a discontinuity of the vortex line 
at $\beta>\beta_c$.  The stabilizing effect of the vortex line tension diminishes as $\epsilon$ decreases due to nonmagnetic impurities or a uniaxial crystal anisotropy, which facilitates the vortex teardown  instability at smaller $\beta$. This brings about the following issue: a curvilinear vortex, as well as any other line topological defects~\cite{mermin}, can hardly break into disconnected pieces, as it would produce large energy barriers growing with the separation between the pieces.  This effect is not incorporated into the force balance Eq. (\ref{dyneq}) which assumes a rigid vortex core and does not ensure the conservation of the topological charge. Thus, Eq. (\ref{dyneq}) only indicates the vortex shape instability because the force balance can no longer be sustained at $\beta=\beta_c$, but it cannot describe the dynamics of the vortex at $\beta>\beta_c$. The latter would require a self-consistent calculation of the spatial distributions of the complex order parameter in the moving vortex core and circulating supercurrents.  

Based on the continuity of the superconducting order parameter and the conservation of the winding number around the vortex core, we can suggest the following picture of the dynamic shape instability. If the velocity of the vortex tip at the surface exceeds $v_0$, the tip does not get disconnected from a slower part of the vortex but turns into a kink along the vortex core, as depicted in Fig. \ref{fig:Fig12}(a).  Dynamics of such kinks could be simulated using the TDGL equations ~ \cite{embon,kramer,kopnin,tdgl1,tdgl2,anl} which describe both the vortex core structure and strong bending distortions of circulating currents around a moving vortex. Yet the derivation of the TDGL equations \cite{kramer,kopnin} disregards the gradient terms in the kinetic equations describing diffusion of nonequilibrium quasiparticles from  the moving vortex core, which is essential for the LO mechanism  \cite{LO}. Thus, the TDGL equations may not be sufficient for a complete description of the dynamic vortex kinks.    

The dynamic kink formation becomes more transparent in layered superconductors, where a vortex perpendicular to the layers is formed by a stack of 2D pancake vortices weakly coupled by interlayer Josephson and magnetic interactions \cite{pv1,pv2,pv3}. This case shown in Fig. \ref{fig:Fig12}(b) also models the LO instability of a vortex piercing a stack of weakly coupled films in a multilayer~\cite{ag_sust}. If the pancake vortex at the surface moves faster than $v_0$, it accelerates because the viscous drag drops and the rest of the restoring force is produced by weak magnetic interactions with other pancake vortices and by the Josephson string \cite{pv2} depicted in  Fig. \ref{fig:Fig12}(b). The Josephson string caused by the Josephson energy between misaligned pancake vortices results in a long-range restoring force $F_J(y_0)$ between the neighboring vortices spaced by $y_0$~ \cite{pv2}:
\begin{gather}
F_J\sim \epsilon_0 y_0/s\Xi,\qquad y_0\lesssim \lambda_J,
\label{F1} \\
F_J\sim \epsilon_0\Xi^{-1/2},\qquad y_0\gtrsim \lambda_J,
\label{F2}
\end{gather}
where $\lambda_J=s\sqrt{\Xi}$, $s$ is the interlayer spacing, $\epsilon_0=\phi_0^2/4\pi\mu_0\lambda^2$, and logarithmic factors $\sim 1$ were disregarded. 

The LO instability of a stack of pancake vortices at high frequencies is considered in Appendix \ref{App}. In this case the Lorentz force driving slow pancake vortices $(v<v_0)$ in the stack is mostly countered by the viscous drag force, resulting in small amplitudes of oscillations $y_n\ll y_{max}\simeq v_0/f $ for each 2D vortex. For  $v_{0}=100$ m/s and $f=10$ GHz, the maximum amplitude $y_{max}\simeq 10$ nm is much smaller than $\lambda=150-400$ nm in cuprates at $T=0$. However, once the velocity of the pancake vortex at the surface $v\sim H\phi_0/\lambda\eta_0$ exceeds $v_0$, it accelerates rapidly so that the magnetic and Josephson restoring forces may stop the runaway vortex at larger distances $y_0\sim \mbox{min}(\lambda,\lambda_J)$. Thus, the amplitudes of pancake vortices at the surface  increase greatly, resulting in a dynamic kink along the 2D vortex stack. 

In a multilayer comprised of superconducting films of thickness $d$ separated by thick dielectric layers which fully suppress the Josephson coupling, the line tension of a stack of 2D vortices only results from their weak magneto-dipole interactions ~\cite{pv3}. The LO instability first occurs for the vortex in the outer film exposed to the applied field if the net Lorentz force $\phi_0H(1-e^{-d/\lambda})$ exceeds the maximum drag force $d\eta_0v_0/2$, that is:
\begin{equation}
H>H_k=\frac{d\eta_0v_0}{2\phi_0(1-e^{-d/\lambda})},\qquad f\gtrsim \frac{\rho_n\xi^2}{\mu_0\lambda^4}.
\label{hk}
\end{equation}   
At $H_k<H<H_ke^{d/\lambda}$, the amplitude $y_1$ of the fast vortex in the outer film increases greatly, while the amplitudes of slow $(v<v_0)$ vortices in other films $(n=2,3,...)$ remain small, $y_n \ll v_0/f$ (see Appendix \ref{App}).  At $H>H_k$ the amplitude of oscillations of the runaway vortex in the outer film is only limited by the weak LO drag and magneto-dipole interaction with other vortices, small vortex mass and a finite rf period. Because the superconducting phase coherence of 2D vortices in different films at $F_J\to 0$ is lost, the Josephson string which provides confinement of 2D vortices disappears. In this case the LO instability at $H>H_k$ causes a true dynamic teardown of a stack of magnetically-coupled 2D vortices in a multilayer.

\section{Discussion}

This work shows that long trapped vortices driven by a strong Meissner RF current can produce a counterintuitive decrease of the surface resistance $R_i(H)$ with the field amplitude which develops as the frequency increases. Such a field-induced microwave reduction of $R_i(H)$ results from interplay of the nonlinear bending elasticity of a vortex and the decrease of the viscous drag with the vortex velocity. Here the bending rigidity of the vortex stabilizes the LO instability characteristic of short vortices in thin films. This effect opens up opportunities for experimental investigations of nonlinear dynamic behaviors of a driven curvilinear vortex, including the formation of dynamic vortex kinks at strong driving forces.  The dynamic behavior of the vortex can be tuned by changing the concentration of nonmagnetic impurities which make the field-induced reduction of $R_i(H)$ more pronounced as the surface gets dirtier.  Because sparse vortices are driven by dissipationless  Meissner currents, the nonlinear dynamics of vortices is masked by heating effects to a much lesser extent than in the conventional dc or pulse transport measurements \cite{mus1980,klein1985,armenio2007,villard2003,grimaldi2008,inst1,inst2,inst3,inst4}.

The decrease of the residual surface resistance $R_i(H)$ with the RF field can contribute to negative $Q(H)$ slopes observed on alloyed Nb cavities \cite{cav1,cav2,cav3,cav4,cav5}.  The vortex mechanism based on the LO decrease of $\eta(v)$ with $v$ proposed in this work is rather different from the decrease of the quasiparticle BCS surface resistance with the RF field  \cite{ag_prl,ag_sust,kg} or the effect of two-level states at the surface \cite{tl}.  Yet our result that trapped vortices could provide a field-induced reduction of $R_i(H)$ which becomes more pronounced at higher frequencies appears consistent with the recent experiment \cite{fnal} which showed that a negative $Q(H)$ slope in  nitrogen-doped Nb cavities becomes stronger as the frequency increases. The LO vortex mechanism is also in agreement with the low-field behavior of $R_s(H)$ observed on Nb cavities ~ \cite{gigi}. The good fit of the theory to the experimental data with reasonable values of $v_0$ and $B_0$ shown in Fig. \ref{fig:Fig3} indicates that the contribution of trapped vortices can indeed be essential.

The LO mechanism of diffusive depletion of nonequilibrium quasiparticles in the moving vortex core was proposed to describe $\eta(v)$ at $T\approx T_c$ where most transport experiments have been were performed. The behavior of $\eta(v)$ at low temperatures $T\ll T_c$ is not well understood as the calculation of $\eta(v)$ at $T\ll T_c$ requires taking into account  complex kinetics of nonequilibrium quasiparticles along with a self-consistent calculation of the order parameters in a moving vortex core ~\cite{kopnin}.  This problem has not been addressed so far, although models of quasiparticle overheating in the vortex core which can result in $\eta(v)$ similar to Eq. (\ref{LO}) have been proposed ~\cite{shklovsk,kunchur,gc}. Transport measurements of $\eta(v)$ in thin films at $T\ll T_c$ and $v\sim v_0$ are masked by overheating produced by fast vortices. By contrast, measurements of $R_i(H)$  in resonant cavities at a low density of trapped vortices controlled by the dc magnetic field $B_0$ can be used to reveal the behavior of $\eta(v)$ and extract the LO critical velocity $v_0$ at $T\ll T_c$. This could be done by fitting the observed $R_i(H)$ with Eq. (\ref{sp}) at low $B_0$ and frequencies $f\ll f_0$ for which heating is greatly reduced.  For instance, the good fit of $R_i(H)$ for a 1.47 GHz Nb cavity shown in Fig. \ref{fig:Fig3} gave $v_0\simeq 30$ m/s at 2 K and $v_0\simeq 35$ m/s at 1.37 K at the trapped field $B_0\simeq 0.7~\mu$T much smaller  than the Earth field.  Yet increasing $B_0$ can reverse the descending field dependence of $R_s(H)=R_{BCS}(T)+R_i(H)$. Indeed, increasing the density of vortices results in stronger RF overheating, causing an increase of the quasiparticle surface resistance $R_{BCS}\propto \exp [-\Delta/T(H,B_0)]$ with $H$ ~ \cite{gurevich2012,tf2} which can overweight the descending $R_i(H)$. Investigation of the extreme vortex dynamics at $10-100$ GHz may require microcavities  ~\cite{yale}. 

Our numerical solutions of the force balance Eq. (\ref{dyneq}) predict a vortex bending instability at large $\beta$ and $\gamma$. In type-II superconductors this instability can give rise to dynamic kinks along an oscillating vortex at fields well below the thermodynamic critical field $H_c$. Since $H_{c1}$ is reduced by nonmagnetic impurities, while $H_c$ is independent of the mean free path, the field range of the shape instability can be expanded by alloying the surface of a superconductor.   
However, a theoretical framework for the description of dynamic vortex kinks which includes the LO mechanism along with a self-consistent calculation  of the vortex core structure and circulating currents around an oscillating curvilinear vortex is lacking.  The TDGL equations ensure the conservation of the winding number, preventing the development of the bending instability into a vortex teardown, but they do not incorporate the LO mechanism of $\eta(v)$. In layered superconductors the situation can be further complicated by Cherenkov radiation of fast oscillating pancake vortices connected by the Josephson strings depicted in Fig. \ref{fig:Fig12} (b). For instance, Cherenkov wakes behind fast vortices can trigger proliferation of vortex-antiivortex pairs in planar Josephson junction arrays \cite{paco} and layered superconductors \cite{ahmad}.

\section{ACKNOWLEDGMENTS}
We are grateful to Gigi Ciovati for discussions and for providing us with the experimental data shown 
in Fig. 3. This work was supported by NSF under Grants PHY 100614-010 and PHY 1734075, and by 
DOE under Grant DE-SC 100387.

\appendix

\section{Derivation of $u(x,t)$ and $\dot{u}(x,t)$ at $\gamma\ll 1$} \label{Ap}
Integration of Eq. (\ref{up}) yields:
\begin{equation}
u(x,t)=\beta_{t}\int_{l}^{x}\frac{(1-e^{-x})dx}{\sqrt{1-\beta_{t}^{2}(1-e^{-x})^{2}}}
\label{a1}
\end{equation}
Taking here $g=1-e^{-x}$, we have:
\begin{equation}
u(x,t)=\beta_{t}\int_{x_{l}}^{x}\frac{gdg}{(1-g)\sqrt{1-\beta_{t}^{2}g^{2}}},
\label{a2}
\end{equation}
where $x_l=1-e^{-l}$. Then we differentiate Eq. (\ref{a2}) with respect to $t$:
\begin{equation}
\dot{u}(x,t)=\dot{\beta}_t\int_{x_{l}}^{x}\frac{gdg}{(1-g)(1-\beta_{t}^{2}g^{2})^{3/2}}
\label{a3}
\end{equation}
Integrations of Eqs. (\ref{a2}) and (\ref{a3}) give: 
\begin{gather}
u=\sin^{-1}s_l-\sin^{-1}s+\frac{(x-l)\beta_t}{\sqrt{1-\beta_{t}^{2}}}+
\nonumber \\
\frac{\beta_{t}}{\sqrt{1-\beta_{t}^{2}}}\ln\frac{1-\beta_{t}s+\sqrt{(1-\beta_{t}^{2})(1-s^{2})}}{1-\beta_t s_l+\sqrt{(1-\beta_t^2)(1-s_l^2)}},
\label{us}
\end{gather}
\begin{gather}
\dot{u}=\frac{\dot{\beta}_t}{1-\beta_{t}^{2}}\biggl[\frac{2-e^{-l}}{\sqrt{1-s_l^{2}}}-\frac{2-e^{-x}}{\sqrt{1-s^2}}+\frac{x-l}{\sqrt{1-\beta_t^2}} + 
\nonumber \\
\frac{1}{\sqrt{1-\beta_{t}^{2}}}\ln\frac{1-\beta_{t}s+\sqrt{(1-\beta_{t}^{2})(1-s^{2})}}{1-\beta_{t}s_l+\sqrt{(1-\beta_{t}^{2})(1-s_l^{2})}}\biggr],
\label{du}
\end{gather}
where $s(x)=\beta_t(1-e^{-x})$ and $s_l=\beta_t(1-e^{-l})$. 

If $\alpha\beta^2\gg 1$ the integration in Eq. (\ref{sp}) yields:
\begin{gather}
p=\frac{1}{\alpha_0\sqrt{1-\beta_t^2}}\biggl[l+
\ln\frac{1-\beta_ts_l+\sqrt{(1-\beta_t^2)(1-s_l^2)}}{1+\sqrt{1-\beta_t^2}}\biggr].
\label{sp1}
\end{gather}

\section{LO instability for a stack of pancake vortices} \label{App}

A dynamic equation for a pancake vortex in the layer closest to the surface can be written in the form 
\begin{equation}
\frac{\eta_{0}v}{1+(v/v_{0})^{2}}+ky=\frac{\phi_{0}H}{\lambda}e^{-x/\lambda}\sin\omega t.
\label{pv1}
\end{equation}
Here the vortex mass is neglected, and the spring constant $k$ results from interaction of vortex pancakes on different layers \cite{pv1,pv2,pv3}:
\begin{equation}
k\sim\epsilon_0\left[\frac{\ln(\lambda/s)}{\lambda^2}+\frac{\ln(\lambda_J/s)}{\lambda_J^2}\right],\quad y\lesssim\mbox{min}(\lambda,\lambda_J),
\label{kk}
\end{equation}
where $\epsilon_0=\phi_{0}^{2}/4\pi\mu_{0}\lambda^{2}.$  The spring term $ky$ in Eq. (\ref{pv1}) is a mean field approximation of a restoring force taking into account both the Josephson interaction of pancakes on the neighboring layers and the long-range magnetic interactions of pancakes on different layers \cite{pv1,pv2,pv3}. In Eq. (\ref{pv1}) this magnetic interaction is approximated by a parabolic magnetic cage potential \cite{pv3} represented by the first term in the brackets in Eq. (\ref{kk}.)  

Consider a high frequency limit $ky_{n}\ll\eta_{0}v_{0}$ in which the amplitudes of oscillations of pancakes are small and the spring term can be neglected. 
Then the solution of the quadratic equation (\ref{pv1}) gives the velocity $v_n(t)$ of a pancake on the $n$-th layer at $x_n=sn$, $n=1,2,3...$: 
\begin{equation}
v_n(t)=\frac{v_0h_n(t)}{1+\sqrt{1-h_n^{2}(t)}},
\label{pv2}
\end{equation}
where $h_n(t)=h_n\sin\omega t$ and $h_n=(2H\phi_{0}/\lambda\eta_{0}v_{0})e^{-sn/\lambda}.$ Integration of Eq. (\ref{pv2}) gives  
the following expression for the time-dependent displacement $y_n(t)=\int_0^tv_n(t)dt$ of the $n$-th pancake during the first half-period $0<\omega t<\pi$:
\begin{gather}
y_n(t)=\frac{v_{0}}{\omega}\bigg[\ln\frac{1+h_n}{(1-h_n^{2}\sin^{2}\omega t)^{1/2}+h_n\cos\omega t}
\nonumber \\
-\frac{1}{h_n}\ln\frac{1+\cos\omega t}{(1-h_n^{2}\sin^{2}\omega t)^{1/2}+\cos\omega t}\bigg]. 
\label{pv3}
\end{gather}
The amplitude $y_n$ of oscillations of the $n-$th pancake is obtained by taking the limit  $\omega t\to\pi$ in Eq. (\ref{pv3}):
\begin{equation}
y_{n}=\frac{v_{0}}{\omega}\left[\ln\frac{1+h_n}{1-h_n}+\frac{1}{h_n}\ln(1-h_n^{2})\right].
\label{pv4}
\end{equation}
The above solutions for $y_n(t)$ exist only below the LO threshold $h_n<1$ on each layer. The LO instability  first occurs on the layer closest to the surface $(n=1)$. Setting $h_1\to 1$ in Eq. (\ref{pv4}), yields the maximum amplitude of the vortex at the LO threshold:
\begin{equation}
y_{max}=\frac{v_{0}}{\pi f}\ln2.
\label{pv5}
\end{equation}
For $v_{0}=10-100$ m/s and $f=1$ GHz, the amplitude $y_{max}\simeq 10-100$ nm is smaller than typical values of $\lambda=150-200$ nm in cuprates at $T=0$. 

The elastic term in Eq. (\ref{pv1}) is negligible if $\eta_{0}v_{0}/2\gg ky_{max}$, that is, 
\begin{equation}
f\gg k/\eta_{0}.
\label{pv6}
\end{equation}
In very anisotropic superconductors (like Bi$_2$Sr$_2$Ca$_2$Cu$_3$O$_{10+\delta}$ or Bi$_2$Sr$_2$CaCu$_2$O$_{8+\delta}$) the Josephson length $\lambda_J\Xi^{1/2}\simeq 700-800$ nm exceeds $\lambda\simeq 200-400$ nm, in which case $\lambda^{-2}\gg \lambda_J^{-2}$ in Eq. (\ref{kk}) and the spring constant $k\sim\epsilon_0/\lambda^2$ at $y\lesssim \lambda$ is mostly determined by the magnetic interaction of vortex pancakes. Assuming that the Bardeen-Stephen $\eta_0$ is applicable, we can re-write the condition  (\ref{pv6}) as follows:  
\begin{equation}
f\gg f_p\sim\rho_n\xi^2/\mu_0\lambda^4.
\label{pv7}
\end{equation}
Taking here $\rho_{n}\simeq1\:\mu\Omega\cdot$m, $\lambda \simeq 200$
nm, $\xi=1.5$ nm for  Bi$_2$Sr$_2$Ca$_2$Cu$_3$O$_{10+\delta}$ at $T=0$, we get $f_{p}\simeq 1$ GHz. Here $f_p(T)$ decreases with $T$ and vanishes at $T_c$. Even though the elastic restoring forces have little effect on $u_n(t)$ at $f\gg f_p$, they nevertheless hold the vortex pancake stack together in the rf field. 

 The LO field threshold $h=1$ occurs at the surface, $2H\phi_{0}/\lambda\eta_{0}v_{0}=1$,
giving the onset of kink formation:
\begin{equation}
B_{k}=\frac{\lambda\eta_{0}v_{0}\mu_{0}}{2\phi_{0}}\simeq\left(\frac{v_{0}\mu_{0}\lambda}{2\rho_{n}}\right)B_{c2}
\label{pv8}
\end{equation}
For $B_{c2}=100$ T, $\lambda=200$ nm, $\rho_{n}=1\,\mu\Omega\cdot$m,
and $v_{0}=1-100$ m/s, we obtain $B_{k}\simeq(10^{-2}-1)$ mT, which
can be below $B_{c1}$ parallel to the ab planes in YBa$_2$Cu$_3$O$_{7-\delta}$.

\end{document}